\documentclass[twocolumn]{aastex631}
\usepackage{relsize}
\usepackage{makecell}
\usepackage{longtable}
\usepackage{supertabular}
\usepackage{graphicx,xcolor} 
\usepackage{booktabs} 
\usepackage{array}    
\usepackage{multirow}
\usepackage{amsmath}
\usepackage{tablefootnote}
\usepackage{comment}
\usepackage{textcomp}
\usepackage{threeparttable}
\usepackage{natbib}

\begin{document}
\title{Can fractal dimension distinguish between grand-design and flocculent spiral arms?}
\author{Biju Saha}
\affiliation{Indian Institute of Science, Education and Research, Tirupati 517619, India}

\author{Subhadip Dutta}
\affiliation{Indian Institute of Science, Education and Research, Tirupati 517619, India}

\author{Arunima Banerjee}
\affiliation{Indian Institute of Science, Education and Research, Tirupati 517619, India}

\correspondingauthor{Biju Saha}
\email{bijusaha930@gmail.com}

\begin{abstract}

About two-thirds of disk galaxies host spiral arms, ranging from well-delineated grand-design spirals to fragmented flocculent spiral galaxies. We introduce fractal dimension $D_B$ as a non-parametric measure to distinguish between grand-designs and flocculents. We calculate the $D_B$ of 197 grand-designs and 322 flocculents from SDSS DR18, using the samples of \citet{Buta..2015} and \citet{Sarkar..2023}. Our calculated median values of $D_B$ are $1.29^{+0.06}_{-0.04}$ and $1.38^{+0.05}_{-0.06}$ for the grand-designs and flocculents, respectively. In addition, a Kolmogorov–Smirnov (K-S) test rejects null hypothesis that these distributions are drawn from the same population. Finally, using a Random Forest (RF) model, we compare the effectiveness of $D_B$ in classifying spiral arm morphology, as compared to five other parameters viz. total atomic hydrogen HI mass $M_{HI}$, ratio of atomic hydrogen mass-to-blue luminosity $M_{HI} /L_B$, concentration index $C_i$, clumpiness $S$ and arm-contrast $C$. Our results indicate that $D_B$ has the highest feature index (30.8\%), followed by $C_i$ (26.0\%) and $M_{HI}$ (21.0\%). In fact, C, the metric routinely used to distinguish between the spiral morphologies has a feature importance of 8.3\%. Further, $D_B$ for grand-designs is found to anti-correlate with the central velocity dispersion with a correlation coefficient of -0.3 and $p \ll 0.05$. A high value of central velocity dispersion indicates a central Q-barrier, which favors the formation of grand designs according to the density wave theory. Thus, fractal dimension serves as a robust metric to distinguish between spiral morphologies and also links to the formation mechanism of spiral features.

\end{abstract}


\section{Introduction}

Almost all massive disk galaxies in the nearby universe host magnificent spiral arms (\citealp{Nair..2010, Willett..2013}). Being a non-axisymmetric feature in the galactic disk, the spiral arms exert gravitational torques on the interstellar matter, driving an outward transport of angular momentum and an inward flow of gas, thus feeding the galactic centre including the active galactic nuclei. (\citealp{LyndenBell..1972}).  Besides, as the spiral arms sweep through the interstellar gas, they shock and compress the same, triggering star formation and serving as the primary sites of star formation in the galactic disk (\citealp{Dobbs..2011}). The spiral arms therefore play a crucial role in regulating the secular evolution of galaxies. Interestingly, they also manifest a wide spectrum of morphologies ranging from the well-delineated grand-designs, to the irregular and patchy flocculants, and sometimes a combination of both. Although several theories have been proposed to model the dynamics of spiral arms, their origin and evolution are still not clearly understood.\\

\citet{Lin..Shu..1964} proposed that spiral arms arise from quasi-stationary density waves in the galactic disk, rotating with a constant pattern speed like a rigid body. In other words, the galactic spirals are normal modes of oscillations of the disk due to a non-axisymmetric perturbation. However, according to density wave theory, these waves get reflected at the Outer Lindblad Resonance (OLR) and propagate inward, eventually getting absorbed at the Inner Lindblad Resonance (ILR) (\citealp{Toomre..1969}). This raises questions about the sustenance of spiral structures which are density waves. To address this, \citet{Mark..1976} proposed the WASER (Wave Amplification by Stimulated Emission of Radiation) mechanism, in which a wave is reflected outward at the Q-barrier before reaching the ILR. At the Q-barrier, the Toomre Q parameter is high enough to inhibit wave propagation. The resulting reflected wave can then travel outward, potentially setting up a standing wave pattern between the Q-barrier and the OLR, thus sustaining the spiral density wave. Bar-driven instabilities or tidal interactions are typically considered to trigger global spiral waves, organizing the interstellar medium into coherent arm structures (\citealp{Toomre..1977, Athanassoula..1980, Elmegreen..1982, Ohe..2015}). In contrast, flocculent spirals display patchy, fragmented arm segments, formed from stochastic star formation driven by local gravitational instabilities (\citealp{Elmegreen..1991,Li..2005,Gerola..1978,Sleath..1995}). These localized instabilities can arise from variations in gas density, turbulence, or feedback from previous generations of stars, leading to a more chaotic and irregular appearance without the presence of large-scale density waves. Previous studies reveal the dark matter (DM) halo also an influence on the development of spiral arms. While the presence of a DM halo suppresses the local spiral instabilities in the disk by increasing the Toomre Q parameter, it has very little impact on the global spiral modes (\citealp{Ghosh..2014, Ghosh..2016}). Further, \citet{Ganes..2024} observed that disk modeled as a gravitationally coupled two-component system of stars and gas embedded in a spherical halo does not produce any spiral features, but the quadrupolar potential of an oblate DM halo facilitates a transient global spiral pattern in the disk. \\

Despite the enhancement of our theoretical understanding of the dynamics of galactic spiral arms, several aspects of its formation and evolution continue to be puzzling. $N$-body + hydrodynamical simulations typically reveal transient recurrent spiral arms originating from swing-amplified noise (\citealp{Sellwood..1984, Bottema..2003, Onghia..2013}), they fail to generate long-standing grand-design patterns in isolated disk galaxies. Such patterns are more successfully reproduced in the presence of a bar, an interacting companion, or a bulge (\citealp{Saha..2016}) i.e., in the presence of a forcing term. In fact, in compliance with above, \citet{Bittner..2017} studied spiral galaxies in the Spitzer Survey of Stellar Structure in Galaxies ($S^4G$) sample and found grand-designs support more massive bulges. However, they also detected prominent spiral arms in a small subset of unbarred sample, which poses questions about the theoretical understanding that the density waves seen in grand-design are bar-driven. Besides, \citet{Kennicutt..1981} found that the pitch angle of the spiral arm correlates more strongly with the maximum rotational velocity than the central mass concentration and Hubble type of the spiral arm, contradicting the density wave theory. Further, \citet{Savchenko..2013} noticed radial variations in pitch angle in a sample of 50 spiral galaxies, a phenomenon not yet understood in terms of density wave theory. Finally, some galaxies are observed to exhibit both flocculent and grand design structures. A robust classification scheme and larger samples are needed to refine theoretical models and constrain simulations across diverse parameter space.\\

\citet{Elmegreen..1982, Elmegreen..1987} devised a twelve-point classification scheme to qualitatively distinguish between the nature of spiral arms based on the regularity of the arm structures. Arm classes (AC) 1-4 were considered as flocculent spirals, because they exhibit ragged and patchy arms. AC 5-12 were designated as grand-design spirals, possessing continuous and symmetric arms. Later, an intermediate class (AC 5-9) was identified as multiple arm spirals, with more than two arms or inner two-arm symmetry branching to multiple long arms. \citet{Elmegreen..2011} applied this morphological classification to the 36 spiral galaxies in the $S^4G$ sample, which was later extended by \citet{Buta..2015} to 1074 spiral galaxies in the same survey. However, the adopted classification scheme of spirals into different classes mentioned in all the above studies was based on visual inspection and is therefore subjective. To quantify the visual classification of spiral arms, \citet{Elmegreen..2011} and later \citet{Bittner..2017} used $S^4G$ images to calculate arm contrasts, finding that grand-designs have higher median contrasts than flocculents. However, the difference in the values was less pronounced than expected from visual classification, with the distribution showing significant overlap between the two classes. \citet{Yu..2020} analyzed 4378 SDSS spirals to study arm strength and pitch angle, suggesting grand-designs generally have smaller pitch angles and fewer arms. They introduced the fractional $m = 3$ Fourier amplitude ($f3$) as a proxy for arm number. Interestingly, \citet{Thanki..2009} found that fractal dimension serves as an efficient tool to distinguish between elliptical and spiral galaxies. Assuming that the degree of irregularity in isophotes of spiral galaxies may distinguish between different spiral morphologies, we propose fractal dimension as an effective parameter to determine the same, in this paper. Further, fractal dimension may also serve as a useful metric to quantify the irregularities arising out of fine features in spirals like feathers and spurs (\citealp{Feather..2006, Feathers..2022, Feathers..2022..part2}), which are expected to become noticeable and reproducible with deeper surveys like JWST (\citealp{JWST..2006}), EUCLID (\citealp{Euclid..2022}), the upcoming CSST (\citealp{CSST..2011}) and simulations with increasing resolutions.\\

The idea of fractal geometry has been explored to address the hierarchical structure formation mechanism across different length scales. \citet{Yadav..2010} applied fractal analysis to galaxy surveys to test cosmic homogeneity. Besides, interstellar gas in galaxies follows a self-similar structure, characterized by a fractal dimension value of $\sim 2.7$ (\citealp{Elmegreen..2004, Sanchez..2005, Sanchez..2007}). Furthermore, the fractal dimension measured from HII regions serves as an indicator of turbulence in the interstellar medium \citep{Ortiz..2015}. The spatial distribution of newly formed stars is also expected to exhibit a fractal nature (\citealp{Elmegreen..2001,Elmegreen..2003,Bastian..2007}). In addition, \citet{Hodge..1985} suggested that the fractal dimension of the distribution of HII regions may differ between grand-design and flocculent galaxies; however, the findings remain inconclusive due to the limited sample size and the fact that HII regions do not reliably trace the underlying spiral structure. Using CCD images of 89 spiral galaxies and 14 elliptical galaxies, \citet{Thanki..2009} showed that the average fractal dimensions are higher for spirals than for ellipticals. Although it remains uncertain whether such fractal patterns show any correlations with any physical properties of the host galaxies, the presence of similar scaling relations implies that comparable physical mechanisms may be shaping these systems. In this study, we calculate the fractal dimension of 322 flocculent spirals and 197 grand-design spirals using galaxy images from the SDSS DR18 i-band, as catalogued by \citet{Buta..2015} and \citet{Sarkar..2023}. In addition to the fractal dimension, we calculate two other morphological parameters: arm/inter-arm contrast $C$ and clumpiness $S$. These two parameters, like the fractal dimension, are non-parametric measures directly derived from CCD images of galaxies. We also obtain total atomic hydrogen HI mass ($M_{HI}$), ratio of atomic hydrogen mass-to-blue luminosity $M_{HI}/L_B$ and concentration index $C_i$, all of which are known to exhibit distinct distributions between the two spiral classes. Finally, we then determine the most efficient parameter in distinguishing flocculent from grand-design spirals using a random forest classifier . \\

The paper is organized as follows. We describe the sample of grand-design and flocculents spirals used in this study in Section \ref{sec:sample_selection} and the methods of obtaining various parameters are described in Section \ref{sec:method_analysis}. We describe the random forest model in Section \ref{sec:random forest}. We present the results in Section \ref{sec:results} and finally the conclusions in \ref{sec:conclusion}.

\section{Sample Selection}
\label{sec:sample_selection}
We utilize galaxy images from SDSS Data Release (DR18), which offers a large sample of nearby galaxies. The improved resolution allows morphological features to be clearly distinguished. To begin with, we consider the SDSS images of grand-design and flocculent galaxies as classified in \citet{Buta..2015}. \citet{Buta..2015} used the arm classification labels introduced by \citet{Elmegreen..2011} to a sample of 1,114 $S^4G$ galaxies, assigning them into three classes, namely grand-design, flocculent, and multiple-arm spirals. Out of the 201 grand-design and 553 flocculent galaxies classified in \citet{Buta..2015}, we obtain the $i$-band (7480 \text{Å}) FITS images of 123 grand-design and 321 flocculents from SDSS DR18; the remaining being outside the SDSS footprint could not be obtained. Further, we visually inspect the images and neglect those that are unclear, or seen to undergo merger, or display ambiguity when compared with the morphological labels of \citet{Buta..2015}. Additionally, we exclude galaxies with $i > 70$\textdegree, since deprojection becomes unreliable for highly inclined galaxies. The inclination angle $i$ and the position angle of the major axis $\phi$ are obtained from the Hyperleda database (\citealp{Makarov..2014}). Finally, we are left with 156 flocculents and 58 grand-designs from the \citet{Buta..2015} sample.\\

\begin{figure}[h]
    \centering
    \begin{minipage}[t]{0.48\textwidth} 
        \centering
        \includegraphics[width=\linewidth]{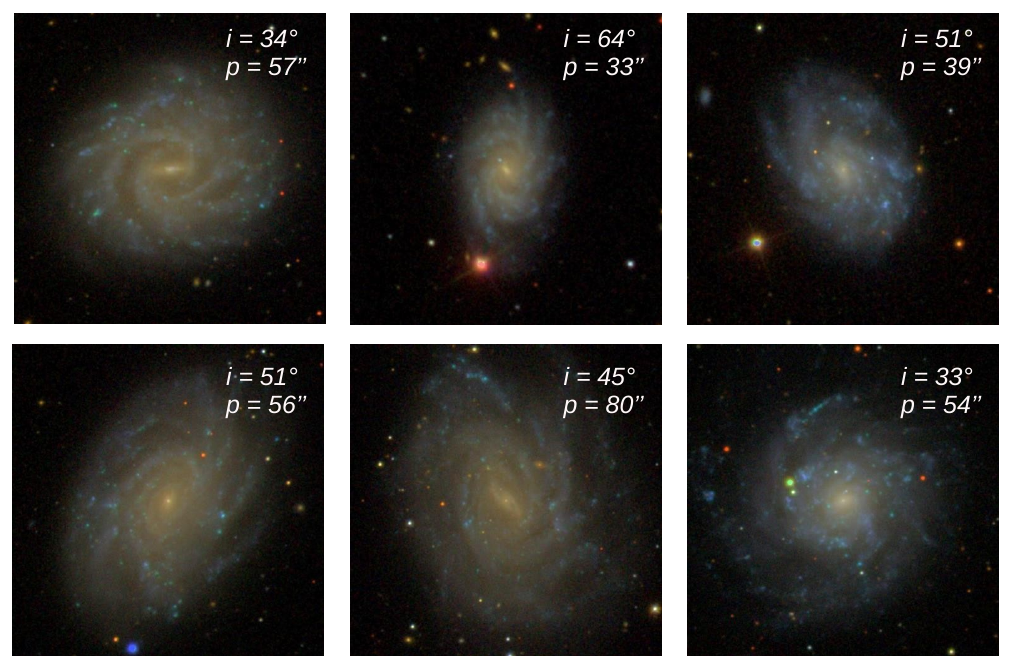} 
    \end{minipage}
    \hfill  
    \begin{minipage}[t]{0.48\textwidth} 
        \centering
        \includegraphics[width=\linewidth]{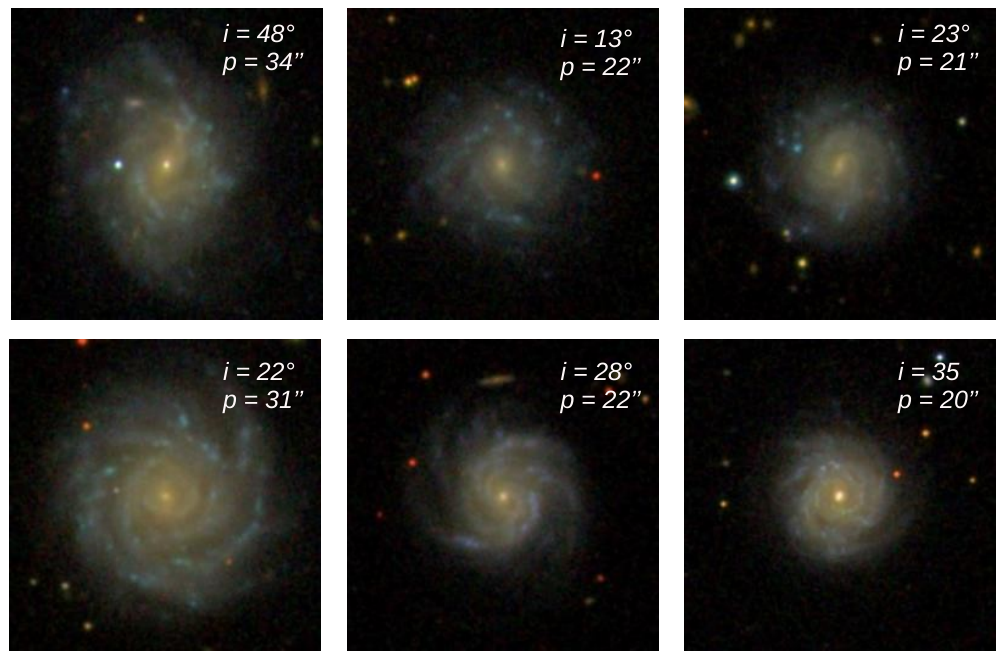} 
    \end{minipage}
    \vspace{-6mm}
    \caption{A subset of the flocculent spirals taken from \citet{Buta..2015} (first two rows) and \citet{Sarkar..2023} (last two rows) from the SDSS DR18, along with the inclination angle $i$ (in degrees; obtained from Hyperleda) and radius containing 90\% of the petrosian flux \textit{petroR90\_i} (in arcsec; collected from SDSS \textit{PhotoObjAll}).}
    \label{fig:flocculent_collage}
\end{figure}

\begin{figure}[h]
    \centering
    \begin{minipage}[t]{0.48\textwidth} 
        \centering
        \includegraphics[width=\linewidth]{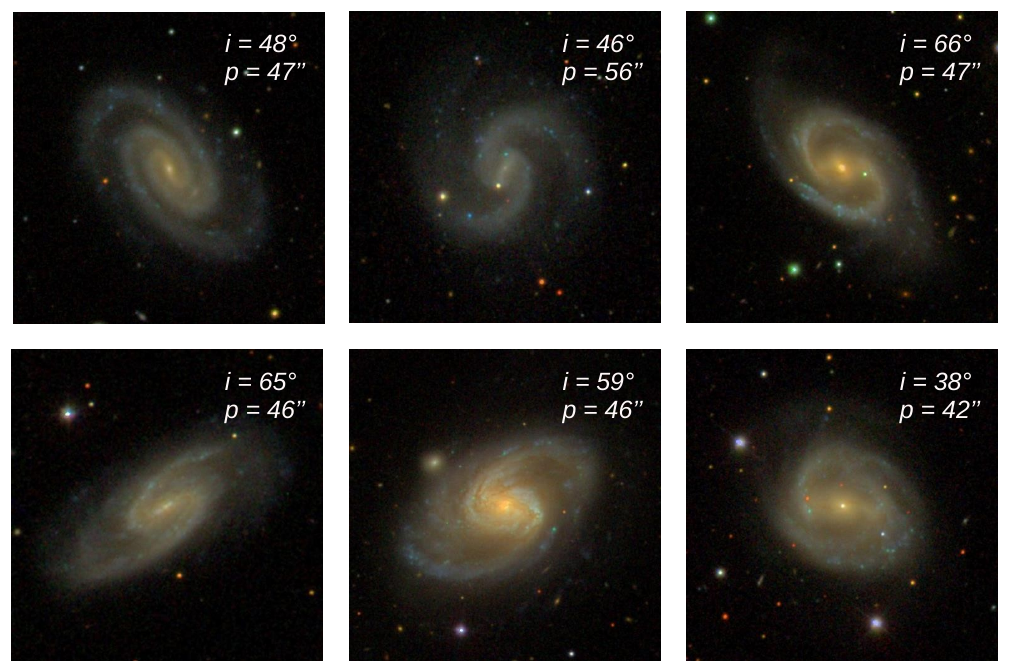} 
    \end{minipage}
    \hfill  
    \begin{minipage}[t]{0.48\textwidth} 
        \centering
        \includegraphics[width=\linewidth]{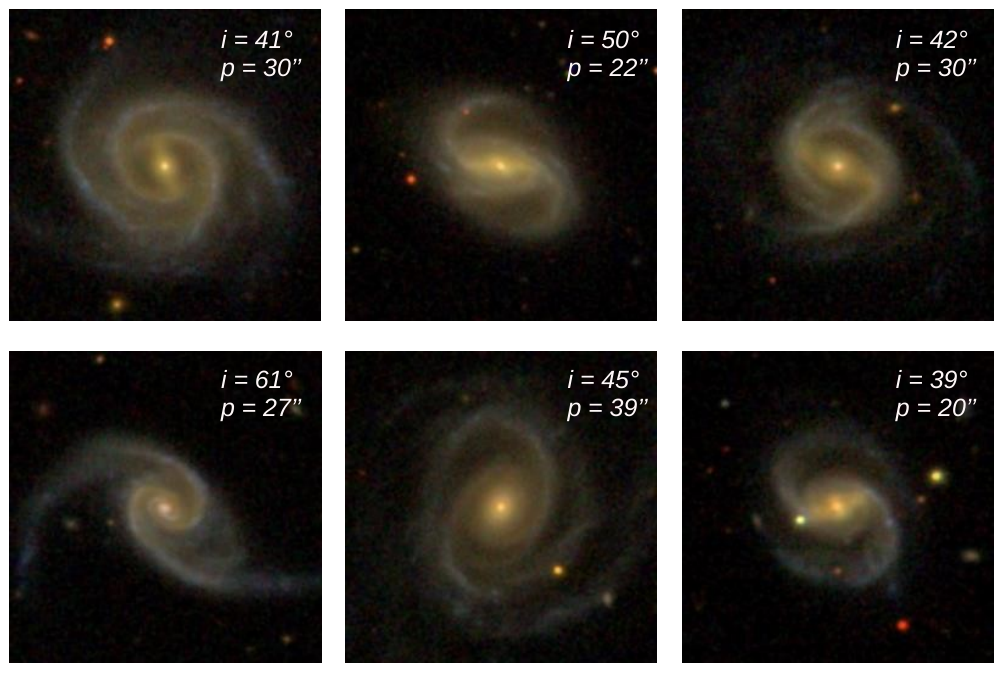} 
    \end{minipage}
    \vspace{-6mm}
    \caption{A subset of the grand-design spirals taken from \citet{Buta..2015} (first two rows) and \citet{Sarkar..2023} (last two rows) from the SDSS DR18. Other notations are same as \ref{fig:flocculent_collage}.}
    \label{fig:grandesign_collage}
\end{figure}

\begin{figure}[h]
    \centering
    \begin{minipage}[t]{0.48\textwidth} 
        \centering
        \includegraphics[width=\linewidth]{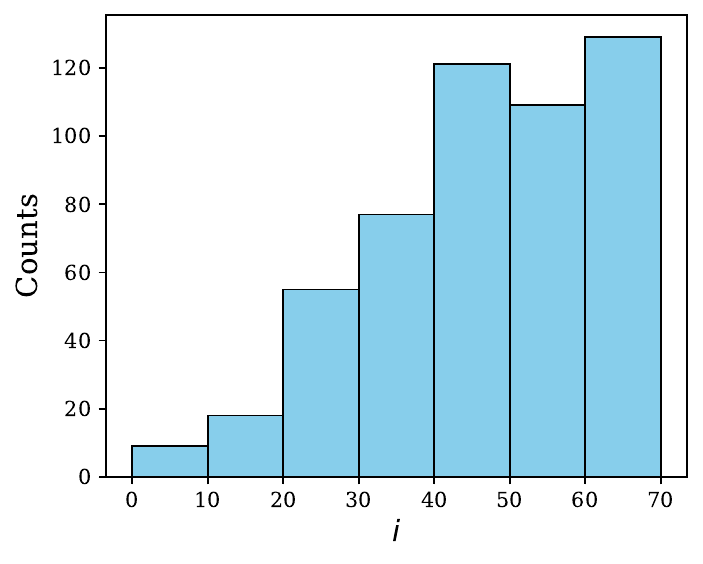} 
    \end{minipage}
    \hfill  
    \begin{minipage}[t]{0.48\textwidth} 
        \centering
        \includegraphics[width=\linewidth]{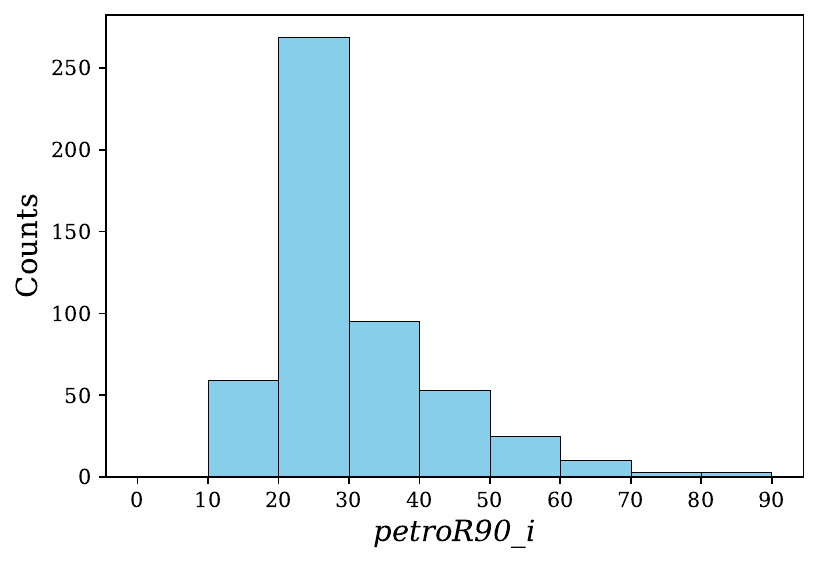} 
    \end{minipage}
    \vspace{-6mm}
    \caption{Distribution of (top) inclination angle \textit{i} (in degrees) and (bottom) Petrosian 90\% light radius \textit{petroR90\_i} (in arcsec) for the galaxies in our sample.}
    \label{fig:inc_petro_hist}
\end{figure}

To augment our sample size, we consider additional grand-design and flocculent galaxies identified by \citet{Sarkar..2023}, who used a Deep Convolutional Neural Network (DCNN) to classify 1,220 SDSS spiral galaxies with an average accuracy of 97\%.  From their catalogue, we selected galaxies classified with a confidence level $>90\%$ and applied the same inclination cut as earlier. This resulted in 166 flocculents and 139 grand-designs from \citet{Sarkar..2023}. To avoid significant class imbalance, we limited the number of flocculents included in the final sample through random selection, ensuring an unbiased representation of both classes. The final dataset (combining samples from both \citealp{Buta..2015} and \citealp{Sarkar..2023}) consists of 322 flocculents and 197 grand-designs. Figure \ref{fig:flocculent_collage} and Figure \ref{fig:grandesign_collage} show a subset of flocculents and grand-designs, respectively; Figure \ref{fig:inc_petro_hist} displays the distributions of the inclination angle $i$ and the Petrosian 90\% light radius $petroR90\_i$; the latter obtained from the SDSS photometric table \textit{PhotoObjAll}.\\

\begin{figure}[h]
    \centering
      \includegraphics[width=\linewidth]{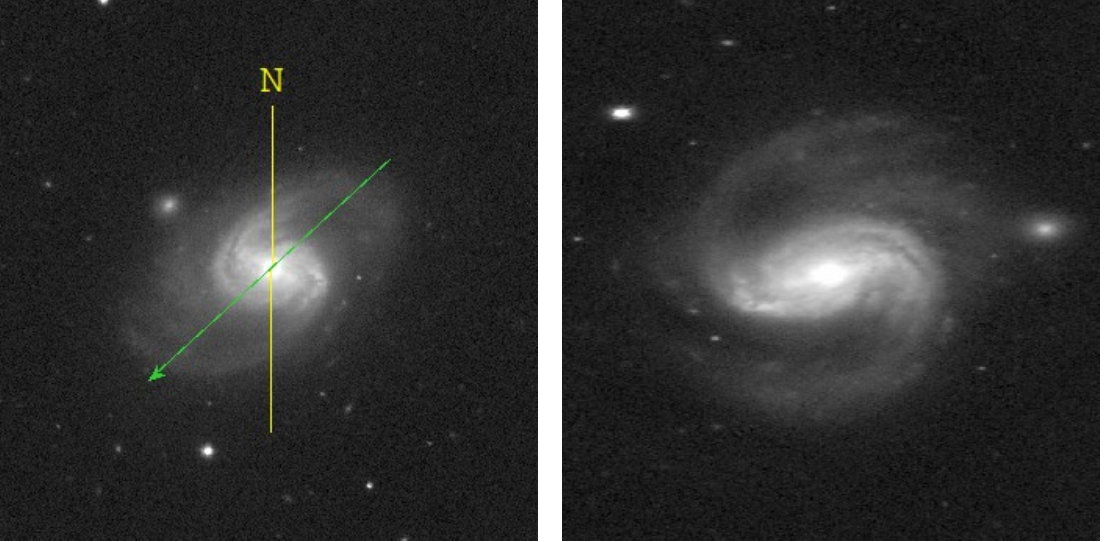}
    \caption{Left: SDSS DR18 i-band (7480 \text{\AA}) image of NGC 3583. NGC 3583 displays a position angle (PA) and inclination angle ($i$) of $126.7^\circ$ and $58.7^\circ$, respectively (values collected from Hyperleda). PA is the orientation of the semi-major axis (indicated by green arrow) in degrees east of north N (indicated by yellow line) and $i$ is the angle made by normal to the plane of galaxy with the line of sight direction. Right: the deprojected image of NGC 3583 obtained by using IRAF task \textit{IMLINTRAN}.}
    \label{fig:deprojected}
\end{figure}

\noindent \textit{Image preparation}: The first step is to obtain a preliminary estimate of the sky. Sky values are determined using the SEP (\citealp{Barbary..2018}) implementation of the SExtractor package (\citealp{Bertin..1996}). The mean sky background is subtracted from each image. Next, an important step in measuring any morphological parameter is to deproject the galaxy to a face-on orientation. For galaxies that are nearly face-on, the plane of the disk parallel to the plane of the sky appears to be circular. A galaxy with an intrinsically circular disk having a random inclination appears on the sky as an ellipse. We  perform deprojection using the inclination angle $i$ and the position angle $\phi$ values. We use the IRAF (\citealp{IRAF}) task IMLINTRAN, which deprojects an image by stretching it along the line of nodes while keeping the total flux conserved (\citealp{Gadotti..2007}). We perform deprojection in order to minimize errors in the resulting measurements of fractal dimension and other parameters, as discussed in the next section. The sky-subtracted, deprojected SDSS $i$-band mage of the grand-design galaxy NGC 3583 is shown in Figure \ref{fig:deprojected}. While the foreground stars do not significantly affect the calculation of fractal dimension and arm contrast (discussed in next sections), it is likely to impact the clumpiness value. Hence, we mask the foreground stars using SEP and mask them by interpolating them with the pixel values chosen from nearby regions.

\section{Morphological classification of spiral arms: Parameters}
\label{sec:method_analysis}

\subsection{Fractal Dimension ($D_B$)}
\label{subsec:fractal_dim}
Mandelbrot introduced the concept of fractals (or self-similar patterns) to describe the irregular complex features observed in nature (\citealp{mandelbrot1983fractal}). While such self-similar patterns are generated by some deterministic rule, fractals that we come across in dynamical systems are usually generated by stochastic processes and, hence do not show any apparent self-similarity (\citealp{Hilborn}). Due to their intrinsic irregularity and rugged appearance, fractal objects do not completely fill the space they occupy —unlike regular Euclidean objects that do fill their embedding space. As a result, fractals are characterized by a dimension that is generally less than the embedding dimension. This leads to the use of fractal dimension, often a non-integer, as a more appropriate measure to describe their geometric complexity.\\ 

One of the most popular method to compute fractal dimension is the box-counting dimension. In this method, the fractal object is overlaid with square boxes, each of size $\epsilon$, and the number of boxes $N$ required to cover the object is noted. The number of boxes $N$ required to cover the object is inversely proportional to the box size $\epsilon$, given by the power law $N \propto \epsilon^{-D_{B}}$, where the exponent $D_B$ is the box-counting dimension. In practice, the fractal dimension $D_B$ in the box-counting algorithm is defined as:\\
\begin{equation}
    D_B=\lim_{\epsilon \to 0} \frac{\log N}{\log 
    {(1/\epsilon)}}
\label{eq:box_counting}
\end{equation}

Irregular appearance in a dynamical system may serve as a probe to the underlying mechanisms. Isophotes, representing contours of constant intensity in a galaxy image, is affected by the active star formation occurring in regions of spiral instabilities. Stochastic star formation, believed to be the dominant mechanism in flocculent galaxies (\citealp{Gerola..1978}), causes the isophotes to appear highly irregular and fragmented. In contrast, grand-design spirals exhibit density waves that concentrate star formation along well-defined spiral arms (\citealp{Lin..Shu..grand}), resulting in a more regular and coherent appearance. The degree of complexity in these isophotes can be quantified using the fractal dimension, which provides a measure of their structural irregularity.\\

\begin{figure}[h]
    \centering
      \includegraphics[width=\linewidth]{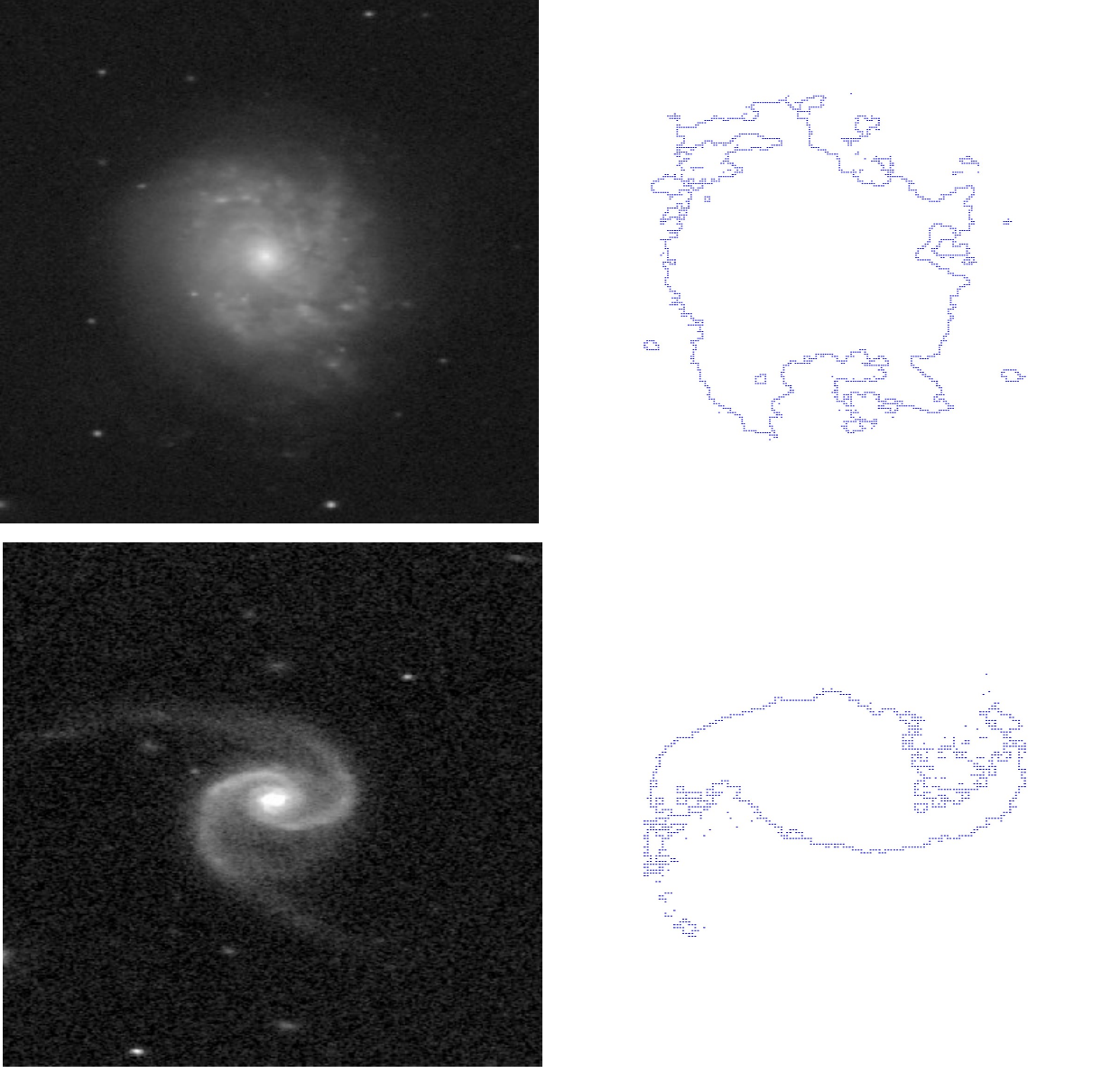}
    \caption{Examples of contours (along with the corresponding SDSS DR18 i-band images) generated at a given intensity level using the method described in \citet{Thanki..2009}. Top row (from left to right): NGC 3949 (a flocculent spiral galaxy) and contour for NGC 3949; Bottom row (from left to right):  J111628.05+291936.1  (a grand-design spiral galaxy); contour for  J111628.05+291936.1. Images are deprojected to face-on orientation before obtaining the contours.}
    \label{fig:fractal_contours}
\end{figure}

The fractal dimension for an isophote corresponding to a given intensity value is calculated by first extracting the contours around the isophote and then applying the method of box-counting dimension. The contour around each isophote is obtained using the methodology prescribed in \citet{Thanki..2009}, which we briefly state below. Each pixel in the galaxy image is assigned a value of 0 or 1, depending on whether the intensity value of the pixel is less than or greater than the given intensity value, respectively. This divides the pixels of the galaxy image into two distinct groups—a set of pixels designated as 0 and another set as 1. The pixels that contribute to the contours around a given intensity value are a subset of pixels designated as 1 but have at least one neighbouring pixels designated as 0. The coordinates of the pixels meeting this criterion are saved to generate an image of the isophotal contour corresponding to the given intensity value. Figure \ref{fig:fractal_contours} shows an example of contours extracted at a given intensity value for a grand-design and a flocculent galaxy from our sample. Clearly, the contour corresponding to the grand-design sample (bottom row) represents the well-defined spiral arms, whereas that of the flocculent sample (top row) represent rugged features.\\

Next, the box-counting algorithm is implemented to obtain the fractal dimension of the contour. For this, the image containing the contour is overlaid with grids of size $\epsilon$, and the number of boxes $N$ which contain the contour is counted. This process is repeated for different grid sizes, ranging from the minimum value of $\epsilon$ where one grid covers one pixel, to the maximum value, where one grid encompasses the entire image. The corresponding $N$ values are recorded. We then fit a straight line to the plot of $\log N$ versus $ \log \epsilon$, and, as per Equation~\ref{eq:box_counting}, the negative slope of the fitted line  represents the fractal dimension of the isophotal contour.\\

For a particular galaxy, we calculate the fractal dimension for a  series of isophotal contours based on a range of intensity values. The intensity range is guided by the radial range within which the spiral arms are prominent. Following \citet{Thanki..2009}, we impose a radial cut between the inner radius $0.25r_P$ and the outer radius $r_P$, where $r_P$ is the Petrosian radius. For the galaxies in our sample, we use the Petrosian 90\% light radius, $petroR90\_i$, as an approximate measure of $r_P$. The inner radius cut is defined to exclude the central regions, which approximately contain the bulge. Near the center, isophotal contours enclose compact regions with fewer pixels. This reduction in data points leads to poor statistical sampling, making the box-counting estimate of the fractal dimension less reliable in these regions. The outer radius cut is chosen to exclude the outer parts of the galaxy where the signal-to-noise ratio becomes low and spiral arms tend to fade. Upon visual inspection, we find that this radial range ($0.25r_P < r < r_P$) is generally suitable for galaxies in our sample. Now, the corresponding intensity range is defined as follows:  The maximum flux $I_{max}$ and minimum flux $I_{min}$ are calculated as the mean flux within an annulus of radius $0.25r_P$ and $r_P$, respectively. The intensity range is equally spaced in log scale so that the outer parts of the galaxy, where the spiral arms dominate, receive more weight when the average fractal dimensions are calculated. The mean of the fractal dimension values obtained over the range of intensity levels represent the overall fractal dimension of the galaxy.\\ 

\noindent \textbf{Effect of foreground stars and inclination:} The CCD images of some galaxies are contaminated by foreground stars. However, the number of additional contours introduced by these stars that could affect the galaxy isophotes is relatively small — and almost negligible for isophotes corresponding to the outer regions. As a result, the derived fractal dimension remains nearly unchanged. To confirm this, we calculate the fractal dimension for the deprojected images both before and after masking the foreground stars and find that the computed values remain consistent up to the second decimal place. In this study, we report the fractal dimension computed before masking to emphasize that masking is not an essential step for computing the fractal dimension. The calculated values for fractal dimension $D_B$ for the complete sample are provided in Appendix \ref{appendix:fractal_values_appendix}.\\

\begin{figure}[h]
    \centering
      \includegraphics[width=\linewidth]{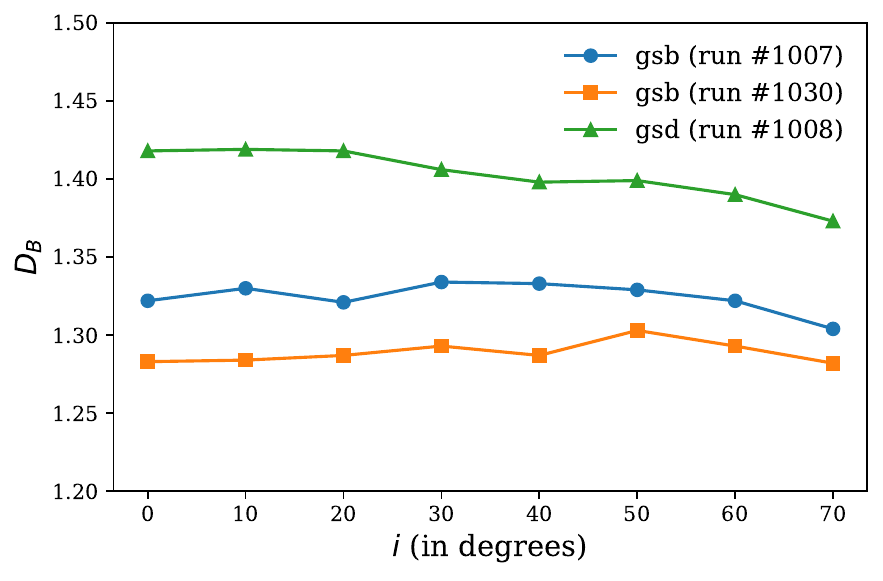}
    \caption{The variations of $D_B$ with $i$ for the three different \textsc{GalMer} models.}
    \label{fig:fractal_inc}
\end{figure}

Furthermore, as noted in \citet{Thanki..2009}, the fractal dimension does not vary significantly with inclination. We verified this using a set of three isolated spiral galaxies from \textsc{GalMer}, a database of N-body + hydrodynamical simulations of galaxy mergers (\citealp{GALMER..2010}). Specifically, we analyzed snapshots from the \textsc{GalMer} models gSb run \#1007, gSb run \#1030, and gSd run \#1008, each taken 150 Myr after the start of the simulation, at inclinations ranging from $i=0$\textdegree  to $i=70$\textdegree (at an interval of 10 \textdegree). To mimic the spatial resolution of SDSS, we convolved each image with a 2D Gaussian kernel with Full-Width at Half-Maximum (FWHM) as 4 pixels, corresponding to the average SDSS Point Spread Function (PSF). We found that the fractal dimension typically decreased at the most by 0.05 ($\approx 3\%$) over this range of inclinations. Figure \ref{fig:fractal_inc} shows the variations of $D_B$ with $i$ for the three different \textsc{GalMer} models. Thus, fractal dimension is an intrinsic morphological feature and does not vary significantly with inclination. 

\begin{figure}[htbp]
    \centering
    \parbox{0.52\textwidth}{
            \centering
            \includegraphics[width=0.8\linewidth]{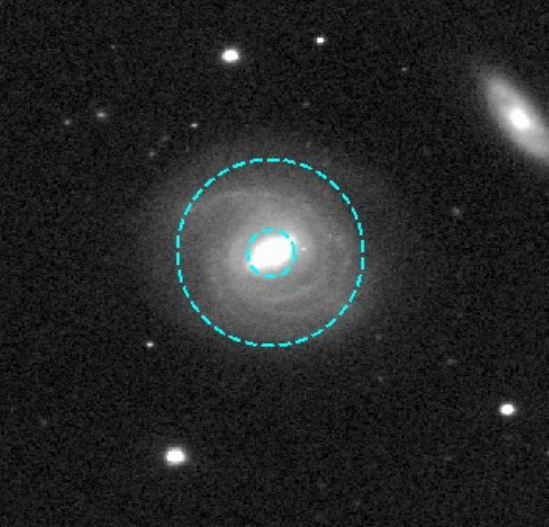}
            \vspace{1ex}

            \includegraphics[width=0.9\linewidth]{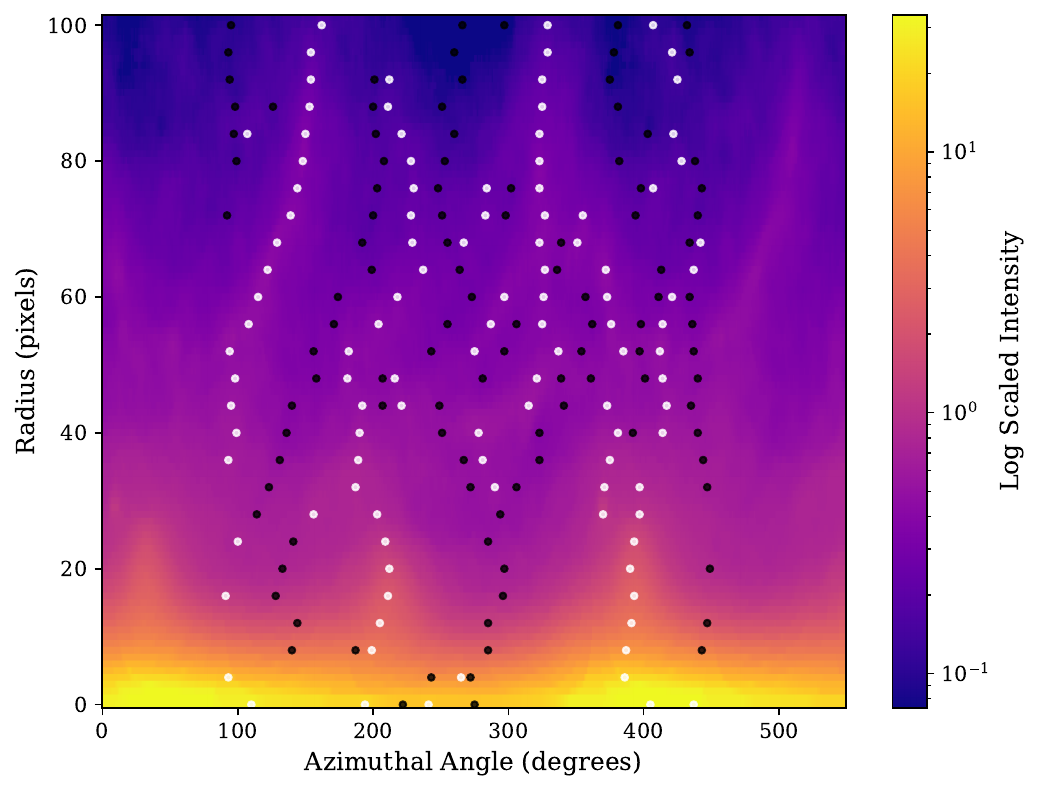}
            \vspace{1ex}

            \includegraphics[width=0.9\linewidth]{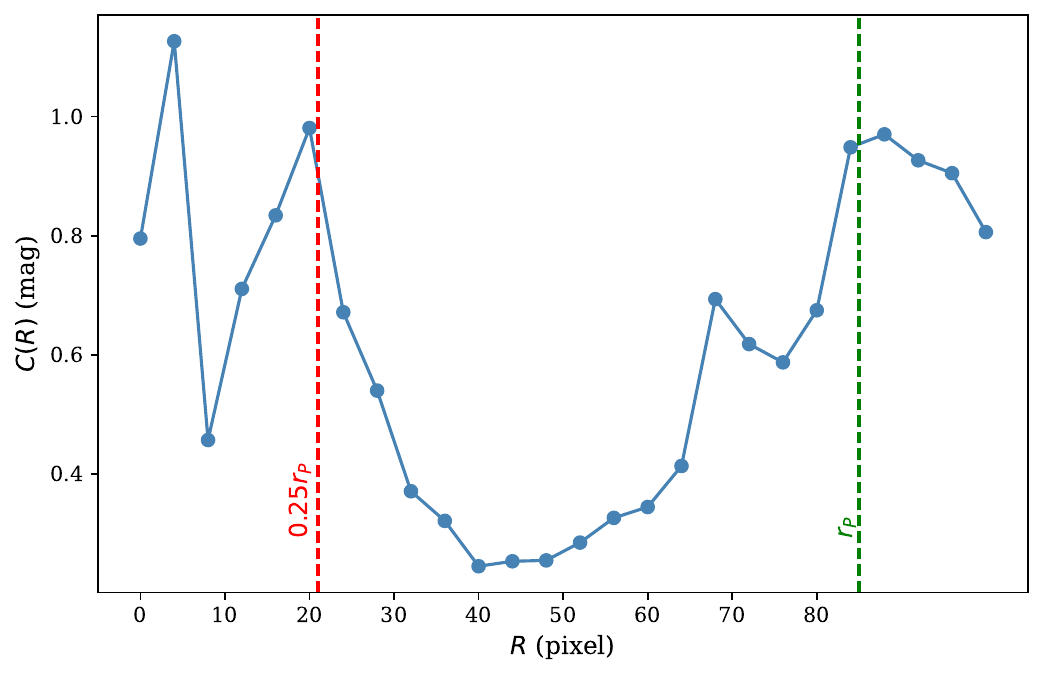}
        }
     \caption{Top: SDSS DR18 i-band image of J094508.97+683540, overlaid with two concentric rings of radius $0.25r_P$ and $r_P$; Middle: Polar plot, with the regions of maxima (white dots) and minima (black dots) denoting arm and inter-arm regions, respectively. Bottom: Plot of arm contrast $C(R)$ versus $R$, with two radial ranges denoted by the dashed line. The median $C$ is calculated within this radial range.}
    \label{fig:arm_contrast}
\end{figure}

\subsection{Arm/inter-arm Contrast (C)}
\label{subsec:contrast}

Spiral arms are sites of active star formation and therefore appear brighter than the surrounding interarm regions. \citet{Elmegreen..2011} introduced the arm/inter-arm contrast as a quantitative measure of this brightness difference, defined as the flux ratio between the spiral arm and the adjacent interarm areas. Grand-design spirals, characterized by well-defined and coherent arm structures, tend to exhibit slightly higher contrast values compared to the more fragmented flocculent arms. \citet{Bittner..2017} performed a detailed measurements of arm/inter-arm luminosity contrasts for almost 290 spiral galaxies galaxies, using the $S^4G$ images. We follow the detailed description outlined in \citet{Bittner..2017} for automating the procedure for the calculation of the arm/inter-arm contrast for the galaxies in our sample of grand-designs and flocculents. Below we briefly describe the procedure adopted from \citet{Bittner..2017}.\\


Although the images are downloaded such that the  galaxy is positioned at the center of the image, we accurately determine the center using IMCENTROID task of the IRAF package. The deprojected face-on images are transformed into polar coordinates, with y-axis representing linear steps of radius and the x-axis is the azimuthal angle from 0 to 360\textdegree. The radial steps are taken to 4 pixels, which is almost equivalent to the the Full-Width at Half-Maximum (FWHM) of the Point Spread Function (PSF) for SDSS (PSF\textsubscript{FWHM} $\sim$ 1.4 arcsec; pixel scale $\sim$ 0.396 arcsec). The azimuthal profile for the intensity are obtained for each radius ranging from $0.25r_P$ to $1r_P$. The presence of any foreground stars show up as a sudden spike in the intensity profile. To remove such kind of outliers in the intensity distribution, we replace it with the median intensity value in rectangle with a width of 10 azimuthal degree and a height of 4 pixels. The intensity curve is then smoothed using a savgol filter. For a given radius, the local maxima (representing the arm region) and the local minima (representing the inter-arm region) are detected numerically. Next, we take the average of all minima ($I_{min}$) and all maxima ($I_{max}$) to calculate the arm/inter-arm contrast $C(R)$ as a function of radius $R$ (\citealp{Elmegreen..2011}),
\begin{equation}
    C(R)=2.5 \times \log \left ( \frac{I_{max}}{I_{min}} \right )
\end{equation}

We quantify the arm-contrast for a galaxy by taking the median of the contrast profile in the same radial range defined earlier ($0.25r_P < r < r_P$). Figure \ref{fig:arm_contrast} shows DS9 view of a grand-design galaxy in log-scale, followed by its polar plot indicating regions of maxima and minima, and the extracted radial profile of the arm contrast $C(R)$. 

\begin{figure*}[t]
    \centering
    \includegraphics[width=\linewidth]{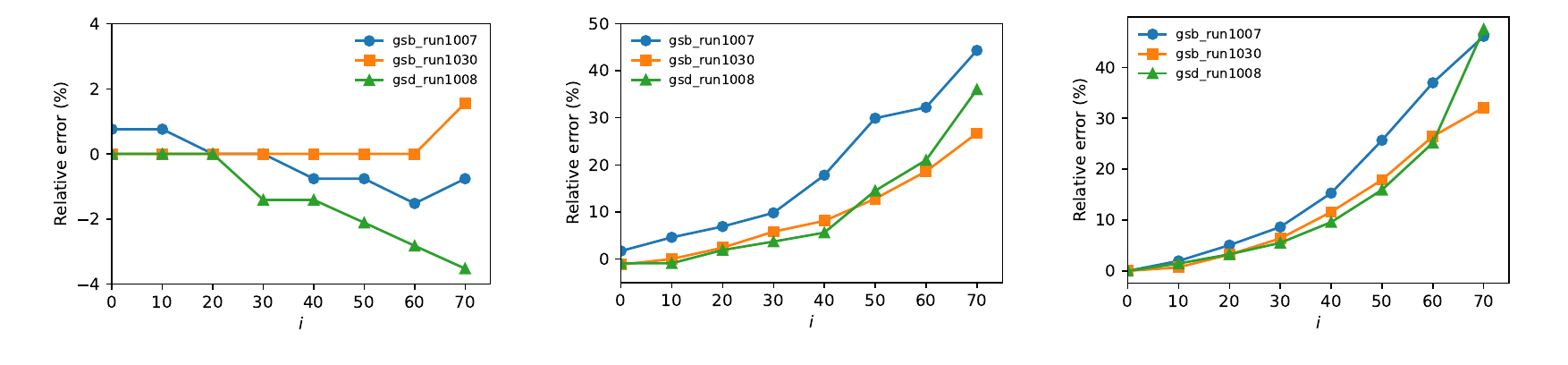}
    \caption{The variation of relative percentage error in the calculation of $D_B$ , $C$ and $S$ (left to right) with inclination $i$, introduced by deprojection. The relative error is calculated as per Equation \ref{equ:relative_error}. }
    \label{fig:error_deprojection}
\end{figure*}

\subsection{Clumpiness ($S$)}
\label{subsec:clumpiness}
\citet{Conselice..2003} introduced three non-parametric, model-independent measures known as the CAS parameters ($C$: concentration, $A$: Asymmetry, $S$: smoothness/clumpiness), which are derived from the structural appearances of galaxies, as seen through their stellar light distributions. They also demonstrate how these parameters correlate with fundamental galaxy properties such as bulge to total light ratios, stellar masses and star formation activity. Clumpiness $S$ measures the patchiness of the light distribution or the small-scale structure, which captures the undergoing star formation, containing large fractions of light at high-frequency. The high frequency component in the light distribution of the galaxy is obtained by subtracting the galaxy image smoothed with a boxcar filter of width $\sigma$ from the original image (\citealp{Lotz..2004}):
\begin{equation}
    S =  \frac{\sum_{x,y} I(x,y)-I_S(x,y)}{\sum_{x,y}I(x,y)} - S_{\mathrm{bgr}},
\end{equation}
where $I(x,y)$ is the sky subtracted flux values of the original image and $I_S(x,y)$ is smoothed image,  which is smoothed by a boxcar of width 0.25$r_p$. The value $S_{\mathrm{bgr}}$ is the average clumpiness of the background. S is summed over the pixels within 1.5$r_p$ of the galaxy’s center. However, because the central regions of most galaxies are highly concentrated, the pixels within a circular aperture equal to the smoothing length 0.25$r_p$ are excluded from the sum. We use the publicly-available statmorph code package (\citealp{statmorph}) to measure the $S$ parameter value for each galaxy in our sample. As clumpiness is likely to get affected by the presence of foreground objects, we use the deprojected, masked images of the galaxies for the calculation of $S$.\\

\noindent \textbf{Deprojection bias:} We also examine the error introduced in the calculation of the three metrics (namely $D_B$, $C$ and $S$) as a result of the deprojection (performed using the IRAF task IMLINTRAN). To do this, we refer to three isolated spiral galaxy simulations from the \textsc{GalMer} simulation suite (as described earlier). For each metric, we calculate the relative percentage error (at a given inclination $i$) introduced due to the deprojection,
\begin{equation}
    \text{Relative error (in \%)}= \frac{m_{i,\text{deprojected}}-m_{i=0}}{m_{i=0}} \times 100
    \label{equ:relative_error}
\end{equation}
Here, $m_{i,\text{deprojected}}$ is the value of a metric $m$ calculated from a simulated galaxy that is intrinsically inclined at an angle $i$ and then deprojected to face-on using IRAF's IMLINTRAN task. The term $m_{i=0}$ refers to the corresponding metric value calculated from an intrinsically face-on image of the simulated galaxy ($i = 0$ \textdegree). $m_{i=0}$ represents the true value, free from any inclination bias. Figure \ref{fig:error_deprojection} shows the variation of the relative error (in \%) with inclination $i$ for $D_B$, $C$ and $S$ (left to right, respectively).\\

Only a small error is introduced in the calculation of $D_B$ (at most $\approx -4 \%$). In contrast, the deprojection error in the measurements of $C$ and $S$ exhibits a monotonically increasing trend with increasing inclination $i$. To avoid introducing significant errors in the calculation of $C$ and $S$, we restrict the calculation of $C$ and $S$ to galaxies with inclinations $i \leq 50^\circ$. This threshold is also consistent with the choice made in previous studies, such as \citet{Bittner..2017}, who adopted the same inclination limit for computing $C$. \\

\noindent \textbf{Physical parameters regulating spiral morphology: } We obtain the total atomic hydrogen HI mass  $M_{HI}$, ratio of atomic hydrogen mass-to-blue luminosity $M_{HI}/L_B$ ($L_B$ is the B-band luminosity) and concentration index $C_i$ for the galaxies with $i \leq 50$ \textdegree. $M_{HI}$ and $M_{HI}/L_B$ are derivable from the observable parameters collected from HyperLeda. Previous studies show that grand-designs have slightly higher $M_{HI}$ and lower $M_{\mathrm{HI}}/L_B$ compared to the flocculents (\citealp{Sarkar..2023, Ghosh..2015}). We use the Petrosian radii (collected from the the SDSS DR18 photometric table) to obtain the concentration index $C_i$ (= $R_{90}$/$R_{50}$), as a proxy for the bulge-to-disk mass ratio. Higher concentration index implies more massive bulge (\citealp{Gadotti..2009}). Grand-designs are observed to have a massive bulge (or a high $C_i$), in order to form a long-standing strong density wave, manifesting as a continuous spiral arm (\citealp{Bertin..1989, Bittner..2017}). Along with previously stated properties (fractal dimension $D_B$, arm contrast $C$ and clumpiness $S$) which are derived from the flux distribution, we include these observable features ($M_{\mathrm{HI}}$, $M_{\mathrm{HI}}/L_B$, $C_i$) described here to perform a comparative studies, as to understand which among them gives a robust tool to classify the two categories of spiral. Table \ref{tab:number_available} lists the number of galaxies (with $i \leq 50$ \textdegree) for which data are available for each of the six parameters; the corresponding values for individual galaxies are provided in Appendix  \ref{appendix:random_forest_appendix}.\\  

\begin{table}[ht]
\centering
\caption{Number of galaxies for which data are available for each of the six parameters. Values of $D_B$ are computed for the total sample of 322 flocculents and 197 grand-designs. For the remaining five parameters, we consider galaxies with $i \leq 50$ \textdegree. However, due to missing entries in HyperLEDA for some galaxies, the $C$, $M_{\mathrm{HI}}$ and $M_{\mathrm{HI}}/L_B$ values could not be retrieved for all galaxies.}
\label{tab:number_available}
\begin{threeparttable}
\begin{tabular}{ccc}
\midrule
\textbf{Parameters} & \textbf{Flocculents} & \textbf{Grand-designs} \\
\midrule
$D_B$\tnote{a} & 322 & 197 \\
$D_B$ ($i \leq 50$ \textdegree)   & 182 & 97 \\
$C$\tnote{a} & 182 & 97 \\
$S$\tnote{a} & 182 & 97 \\
$M_{\mathrm{HI}}$\tnote{b} & 135 & 54 \\
$M_{\mathrm{HI}}/L_B$\tnote{b} & 134 & 51 \\
$C_i$ \tnote{c} & 182 & 97 \\
\bottomrule
\end{tabular}
\begin{tablenotes}
\small
\item[a] Derived from this study.
\item[b] Derived from HyperLeda.
\item[c] Obtained from SDSS DR18 photometric table.
\end{tablenotes}
\end{threeparttable}
\end{table}

\section{Random Forest}
\label{sec:random forest}
All the six parameters or features discussed above, places the galaxies in the sample as a data point in a six-dimensional space, where each axis represents the values of each of the six features. To find an optimal boundary that separates the two classes in our sample (namely, grand-design and flocculents) within this high-dimensional feature space, machine learning methods are essential. Random Forest, a supervised machine learning algorithm, has been widely used in astronomy for both classification and regression tasks. It has been applied in star–galaxy-quasar classification (\citealp{Clarke..2020}), photometric redshift estimation (\citealp{Carlies..2010}), detection of variable stars (\citealp{Richards..2011}), classification of supernovae (\citealp{Lochner..2016}), and even in identifying exoplanet candidates (\citealp{McCauliff..2015}).  In this work, we use the \texttt{Scikit-Learn} (\citealp{Pedregosa..2011}) module’s RF implementation, \texttt{RandomForestClassifier}. \\

\begin{figure}[h]
    \centering
    \includegraphics[width=\linewidth]{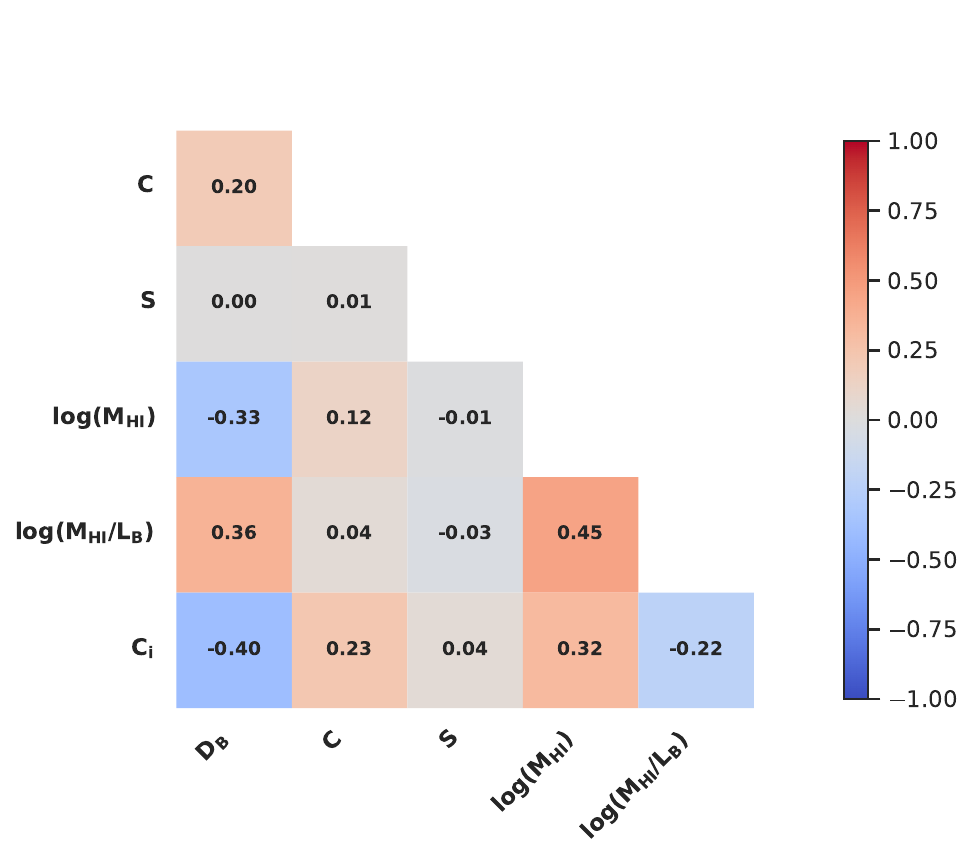}
    \caption{The Pearson correlation coefficient values calculated for each pair of the six features, each of which may distinguish between spiral arm morphology.}
    \label{fig:heatmap}
\end{figure}

Before using the data to train the model, we perform some pre-processing. From the numbers listed in Table \ref{tab:number_available}, we find that approximately 26\% of the flocculent sample and 44\% of the grand-design sample have missing values (denoted as `-' in Appendix \ref{appendix:random_forest_appendix}) for $M_{\mathrm{HI}}$ and $M_{\mathrm{HI}}/L_B$. Excluding these missing entries would result in a significant reduction of the sample size, thus we employ oversampling algorithms to fill the gaps. We applied the \texttt{IterativeImputer} technique (Pedregosa et al. 2011) from the \texttt{Scikit-Learn} library in Python to handle the missing values. The Iterative Imputer method estimates the missing values of a feature using the remaining features as predictors, similar to the approach used in regression models. Next, we also check if the chosen features are showing some correlations because correlated features can introduce unnecessary randomness into the model and bias into determining the importance of various features. Figure \ref{fig:heatmap} shows the heatmap of the correlation matrix, where each element represents the Pearson correlation coefficient ($r$) between two features. All the selected features exhibit little to no correlation ($r < 0.5$). Subsequently, we split the data into training and testing sets using the standard 80:20 ratio. The details of the number of samples used for training and testing are provided in Table \ref{tab:dataset}.

\begin{table}[ht]
\centering
\caption{Sample size for the training and testing datasets.}
\label{tab:dataset}
\begin{tabular}{ccc}
\midrule
 & \textbf{Flocculents} & \textbf{Grand-designs} \\
\midrule
Training & 146 & 77 \\
Testing & 36 & 20 \\
Total & 182 & 97 \\ 
\bottomrule
\end{tabular}
\end{table}

\begin{figure}[h]
    \centering
    \includegraphics[width=\linewidth]{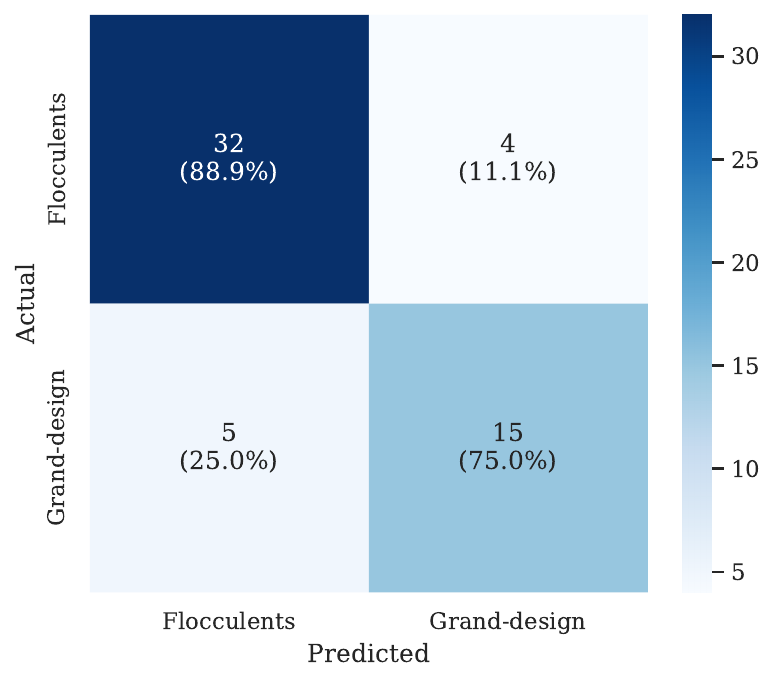}
    \caption{The confusion matrix representing the correct predictions (diagonal elements) and the false predictions (off-diagonal elements) for galaxies in the testing sample. Our trained model achieves a testing accuracy of $\approx$ 84\%.}
    \label{fig:confusionmatrix}
\end{figure}

The performance of a model is examined by how well a trained model performs in classifying an unseen or test dataset. We choose accuracy as the metric for performance of the model. The performance of a model can be controlled by tuning the hyperparameters. We implemented the \texttt{Scikit-Learn} module of \texttt{RandomizedSearchCV} to obtain the best set of parameters for our model. The function was supplied with a dictionary of parameter distributions, which defines the range of values that hyperparameters like number of trees \texttt{n\_estimators} and depth of each decision tress \texttt{max\_depth} is allowed to vary over during the search process. We set \texttt{n\_iter=15}, which means the algorithm samples 15 different combinations of parameters randomly from the defined range. For cross-validation, we used the \texttt{RepeatedStratifiedKFold} strategy with \texttt{n\_splits = 5} and \texttt{n\_repeats = 10}, which are commonly used default settings. The evaluation metric was set to `accuracy', which guided the selection of the optimal hyperparameters based on performance over the validation folds. Table \ref{tab:hyper-param} gives the best set of hyperparameters used for training the model. The best trained model achieved an accuracy of 84\% when applied to the test dataset. Figure \ref{fig:confusionmatrix} gives the confusion matrix, where diagonal elements represents the correctly classified samples (True Positives $TP$ and True Negatives $TN$) and off-diagonal elements represents the false classifications (False Positives $FP$ and False Negatives $FN$). Table \ref{tab:metric} gives the performance metric in terms of precision, recall, $f_1$ score and accuracy. \\

\begin{table}[h]
\centering
\caption{Best set of hyperparameters for the Random Forest model, obtained using the \texttt{Scikit-Learn} module \texttt{RandomizedSearchCV}.}
\begin{tabular}{c c c}
\toprule
& \textbf{Hyper-parameters}& \textbf{Values}\\
\midrule
1 & \texttt{max\_depth} & 3\\
2 & \texttt{n\_estimators} & 87\\
3 & \texttt{max\_features} & \texttt{sqrt}\\
4 & \texttt{scoring} & \texttt{accuracy}\\
\bottomrule
\end{tabular}
\label{tab:hyper-param}
\end{table}

\begin{table}[h]
\centering
\caption{The performance metric of the trained RF model on the testing dataset.}
\begin{tabular}{c c c c}
\toprule
& \textbf{Precision}& \textbf{Recall}& \textbf{$f_1$ score} \\
\midrule
Flocculents & 0.86 & 0.89 & 0.88 \\
Grand-designs & 0.79 & 0.75 & 0.77  \\
\textbf{Accuracy} &  \multicolumn{3}{c}{\textbf{84\%}} \\
\bottomrule
\end{tabular}
\label{tab:metric}
\end{table}

Besides serving as a powerful tool for classification tasks, the Random Forest (RF) algorithm also provides an estimate of the relative importance of each input feature. This is accessed through the default feature importance attribute of \texttt{scikit-learn}'s \texttt{RandomForestClassifier}. Feature importance in RF is determined based on the reduction in Gini impurity contributed by each feature across all the trees in the forest — features that result in greater impurity reduction are assigned higher importance. A discussion on the resulting feature importances and their implications is provided in Section \ref{subsec:results_feature}.

\section{Results \& Discussion}
\label{sec:results}
\subsection{Fractal dimension of flocculents and grand-designs}
\begin{figure}[ht]
    \centering
    \begin{minipage}[t]{0.48\textwidth} 
        \centering
        \includegraphics[width=\linewidth]{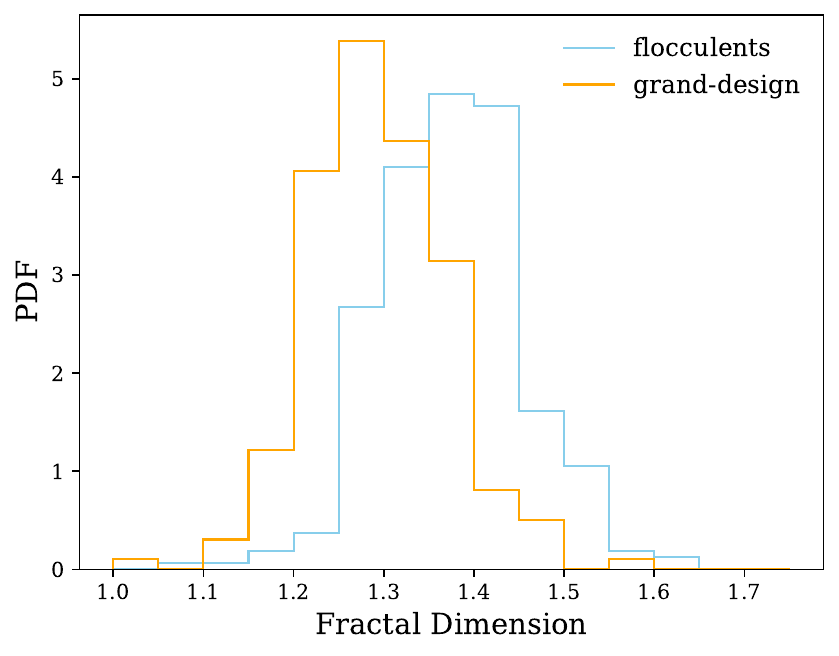}
    \end{minipage}
    \begin{minipage}[t]{0.48\textwidth}  
        \centering
        \includegraphics[width=\linewidth]{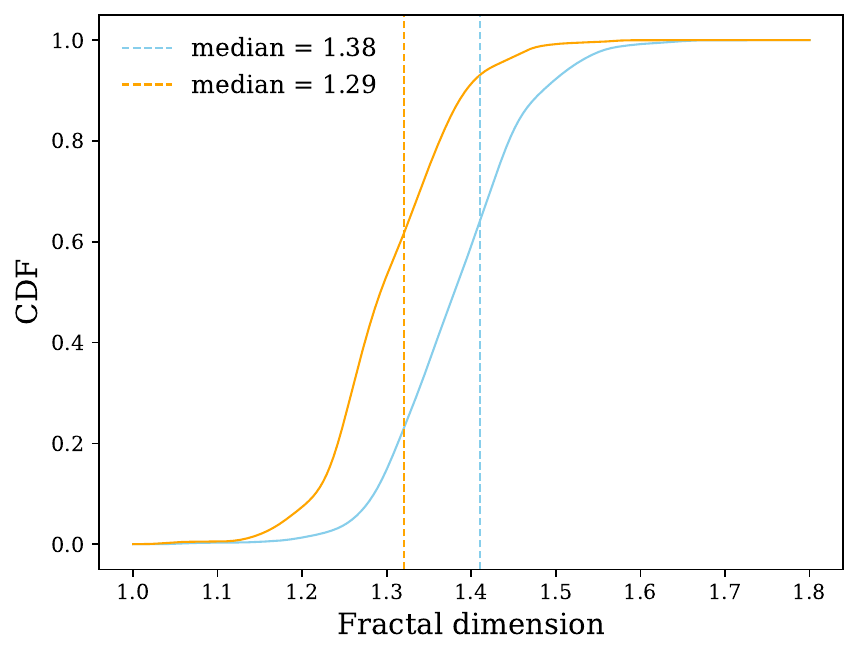}
    \end{minipage}
    \caption{The top panel shows the probability distribution functions (PDFs) for the fractal dimension for the flocculent and grand-design galaxies. The bottom panel shows the corresponding cumulative distribution functions (CDFs).}
    \label{fig:DF_fractal}
\end{figure}

We calculate the fractal dimension using the box-counting method described in section \ref{subsec:fractal_dim}. Figure \ref{fig:DF_fractal} shows the probability distribution functions (PDFs, top panel) and cumulative distribution functions (CDFs, bottom panel) of the fractal dimension for 322 flocculent and 197 grand-design spiral galaxies in our sample. A clear distinction is evident between the distributions corresponding to the two different spiral morphologies. We find the median fractal dimension values to be $1.38_{-0.06}^{+0.05}$ for flocculents and $1.29_{-0.04}^{+0.06}$ for grand-designs, where the superscripts (subscripts) denote the difference between the third quantile (first quantile) and the median value. The irregularities in the fragmented appearance for the flocculents are captured by their relatively higher fractal dimension value. Additionally, a two-sample Kolmogorov-Smirnov (K-S) test was conducted on the fractal dimension values obtained from the flocculent and grand-design galaxy samples. The K-S yields a test statistic of $\Delta_{KS}$ = 0.4110 and a $p$-value $ \ll 0.05$, even at the 100\% confidence level. This suggests a strong rejection of null hypothesis, confirming that the two distributions are significantly different. The results of the K-S test are summarized in Table \ref{tab:KStest}.\\

\noindent \textbf{Discussion based on the sample differences:} In this study, we have considered galaxies from two distinct catalogues. \citet{Buta..2015} consists of low redshift galaxies (median $z$ = 0.005), while the \citet{Sarkar..2023} sample consists of relatively higher redshift galaxies (median $z$ = 0.02). Due to this redshift difference, galaxies in the \citet{Sarkar..2023} sample appear fainter and fine structural features are not well-resolved. In contrast, the lower-redshift \citet{Buta..2015} sample are well-resolved. This difference in image resolution likely contributes to the systematically higher fractal dimension values observed in both grand-design and flocculent spirals from the \citet{Buta..2015} catalogue. The median value of $D_B$ based on the two distinct catalogue is presented in Table \ref{tab:sample_diff}.\\

\noindent \textbf{Discussion on the values of $D_B$ for multi-armed spirals:} 
We also compute the fractal dimension $D_B$ for 111 multi-armed spiral galaxies selected from \citet{Buta..2015}, using the same selection criteria described in Section~\ref{sec:sample_selection}. The median value of $D_B$ for the multi-armed spirals is $1.34_{-0.04}^{+0.04}$. A comparison of the median $D_B$ values for the three spiral types in the \citet{Buta..2015} sample is presented in Table~\ref{tab:sample_diff}. The multi-armed spirals (often considered an intermediate category) might be anticipated to exhibit a $D_B$ value between those of grand-design and flocculent spirals. However, based on our sample, the median $D_B$ value for multi-armed spirals is lower than that of both flocculent and grand-design spirals. It is important to note that in our sample derived from \citet{Buta..2015}, the grand-design spiral class consists of smallest number of galaxies (58), compared to 156 flocculents and 111 multi-armed spirals. A more robust comparison would require a larger and more balanced sample across all three spiral types. We plan to address a detailed discussion on the multi-armed spirals in a future study.

\begin{nolinenumbers}
\begin{table*}[ht]
\centering
\small 
\setlength{\tabcolsep}{10pt}
\caption{The K-S test statistics for the distribution of fractal dimension for our sample of flocculent and grand-design spirals.}
\begin{tabular}{c c c c c c c c}
\toprule
 & \textbf{{Count}} & \textbf{Median} & \textbf{$\Delta_{KS}$} & \textbf{p-value} & \multicolumn{3}{c}{\textbf{$\Delta_c$($\alpha$)}} \\
 & & & & & \textbf{90\%} & \textbf{99\%} & \textbf{100\%} \\
\midrule
\vspace{2mm}
Flocculents & 322 & $1.38_{-0.06}^{+0.05}$ & \multirow{2}{*}{0.4110} & \multirow{2}{*}{ $10^{-19}$} & \multirow{2}{*}{0.0099} & \multirow{2}{*}{0.0132} & \multirow{2}{*}{0.0905}\\ 
Grand-designs & 197 & $1.29_{-0.04}^{+0.06}$ & & & & & \\
\bottomrule
\end{tabular}
\label{tab:KStest}
\end{table*}
\end{nolinenumbers}

\begin{nolinenumbers}
\begin{table}[h]
\centering
\caption{The median values of $D_B$ calculated for flocculents, grand-design and multi-armed spirals for \cite{Buta..2015} and \cite{Sarkar..2023}. The bracketed values indicate the number of galaxies corresponding to the concerned set.}
\begin{tabular}{c c c c}
\toprule
\textbf{Class}& \textbf{\citet{Buta..2015}}& \textbf{\citet{Sarkar..2023}} \\
\midrule
\vspace{3mm}
Flocculents & $1.42_{-0.04}^{+0.04}$ (156) & $1.34_{-0.03}^{+0.05}$ (166) \\
\vspace{3mm}
Grand-design & $1.37_{-0.05}^{+0.03}$ (58) & $1.27_{-0.02}^{+0.05}$ (139)  \\
\vspace{3mm}
Multi-armed & $1.34_{-0.04}^{+0.04}$ (111) & -\\
\bottomrule
\end{tabular}
\label{tab:sample_diff}
\end{table}
\end{nolinenumbers}

\begin{figure*}[ht]
    \centering
    \includegraphics[width=\linewidth]{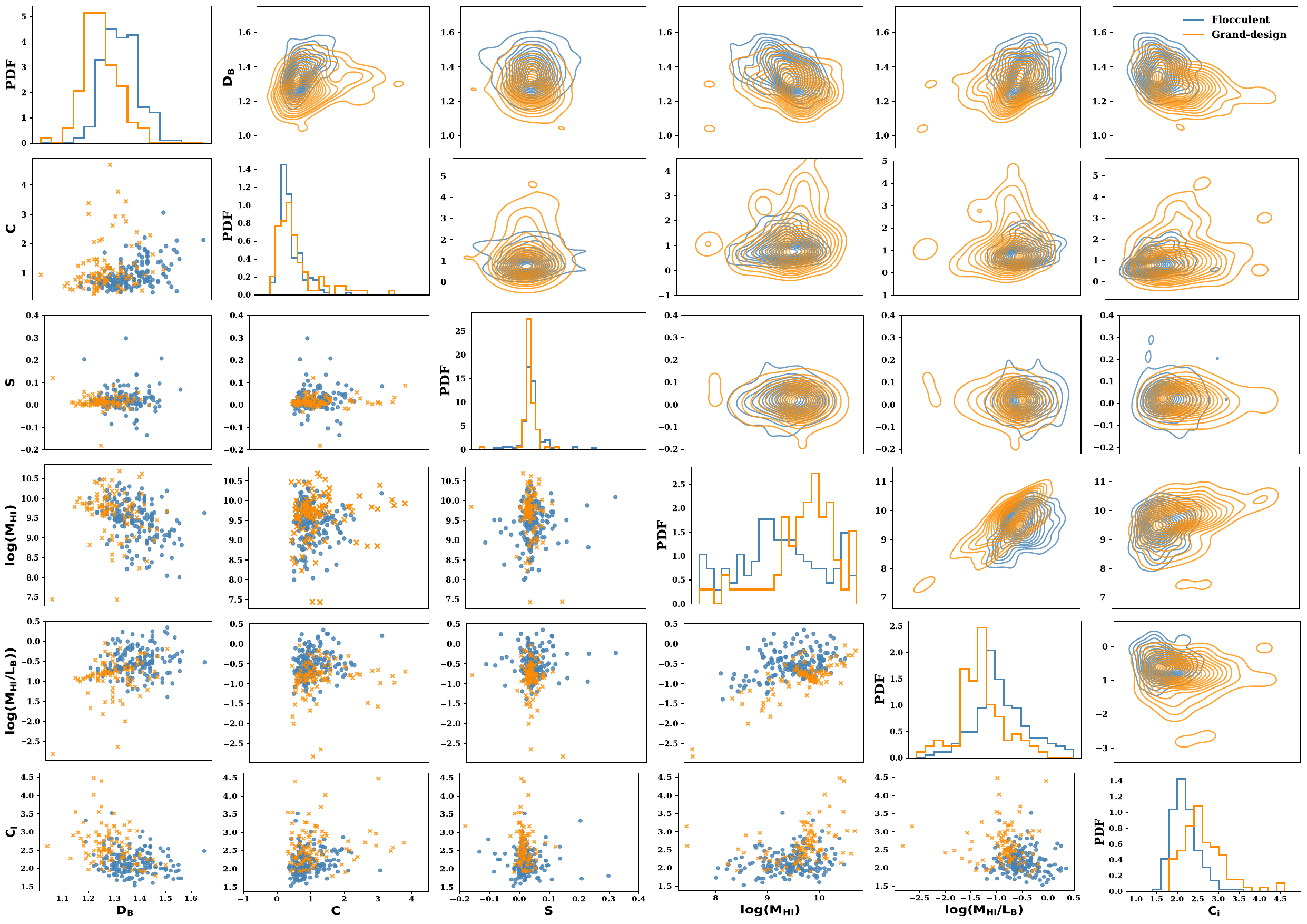}
    \caption{Probability Distribution functions (PDFs) for the six parameters $D_B$, $C$, $S$, $M_{HI}$, $M_{HI}/L_B$, $C_i$ are shown in the diagonal panels. The lower triangular panels show the scatter plots for different pairs of the above six parameters, while the upper triangular panels show the contours of 2–dimensional kernel density estimates (KDEs). The distributions correspond to the sample with $i \leq 50$ \textdegree, which is also used for the random forest analysis.}
    \label{fig:scatter_plot}
    \label{fig:EDA}
\end{figure*}

\subsection{Comparison with other physical parameters}
\label{subsec:results_feature}
Besides the fractal dimension, we also analyze the distributions of five additional parameters: arm/inter-arm contrast $C$, clumpiness $S$, HI mass $M_{HI}$, HI to blue luminosity ratio $M_{HI}/L_B$, and concentration index $C_i$. The median values of these parameters are presented in Table \ref{tab:stat}, and their PDFs are shown in the diagonal panels of Figure \ref{fig:EDA}. Note that for the random forest analysis, we have considered the galaxies which has $i \leq 50$ \textdegree. Grand-designs show slightly higher median values of arm/inter-arm contrast $ C$ compared to flocculents, consistent with earlier studies discussed in Section \ref{subsec:contrast}. Next, flocculents show slightly higher clumpiness $S$, reflecting their inherent patchier appearance. However, there is considerable overlap in the $S$ distributions between two morphological classes, with minimal distinction. In contrast, the fractal dimension, also a non-parametric measure derived directly from galaxy images, similar to $C$ and $S$, proves to be a more effective discriminator between the two morphologies. The distributions of $M_{HI}$, $M_{HI}/L_B$ and $C_i$ for the two classes are also consistent with the trends reported in literature discussed earlier. The lower triangular panels of Figure \ref{fig:EDA} show the scatter plots of each pairwise combination of the six parameters, while the contours of corresponding 2–dimensional kernel density estimates (KDEs) are shown on the upper triangular panels. Among the six features, $D_B$ and $C_i$, individually and jointly yield the strongest separation between the two morphological types. This is followed by $M_{HI}$. Next, we validate these findings through the RF analysis discussed in Section \ref{sec:random forest}. \\

\begin{figure}[ht]
    \centering
    \includegraphics[width=\linewidth]{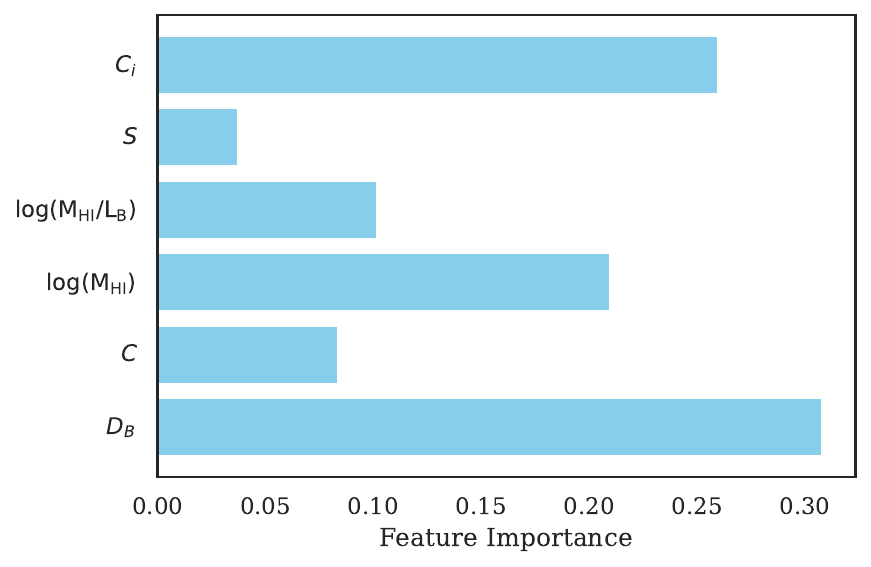}
    \vspace{-7mm}
    \caption{Relative importance of the six features in differentiating between flocculent and grand-design spirals by the random forest algorithm.}
    \label{fig:feature_percentage}
\end{figure}

RF, being a powerful statistical analysis tool, allows us to pin down the feature, out of the lot, which is the most relevant for the classification task. We analyze the feature importance scores from the trained random forest classifier described in Section \ref{sec:random forest}. These importance values provide a relative measure of each feature's effectiveness in distinguishing between the two spiral morphologies. Figure \ref{fig:feature_percentage} shows the feature importance scores for all six parameters based on best-performing model. Fractal dimension $D_B$ have the largest importance (30.8\%) in the random forest classifier, followed by concentration index $C_i$ (26.0\%) and HI mass $M_{HI}$ (21.0\%), in agreement with the interpretations from our previous qualitative analysis. Among the remaining features, gas mass to blue luminosity ratio $M_{HI}/L_{B}$ gains 10.2\%, while arm contrast $C$ and clumpiness $S$ gains little to no importance (8.3\% and 3.7\%, respectively). These findings, combined with results from the K-S test statistics discussed in the previous section, strengthen the argument that fractal dimension—being a non-parametric, model-independent metric derived directly from galaxy images—is a more powerful discriminator between flocculent and grand-design spirals than other measures, including arm/inter-arm contrast.

\begin{nolinenumbers}
\begin{table}[h]
\centering
\caption{Median values of the six features which may distinguish between spiral arm morphology as considered in this study. The values are reported for the sample with $i \leq 50$ \textdegree, which is also used for the random forest analysis.}
\begin{threeparttable}
\begin{tabular}{ccc}
\midrule
 & \textbf{Flocculents} & \textbf{Grand-designs} \\
\midrule
\vspace{5mm}
$D_B$ & \parbox[c]{1.5cm}{$1.36_{-0.05}^{+0.07}$} & \parbox[c]{1.5cm}  {$1.28_{-0.04}^{+0.06}$}\\
\vspace{5mm}
$C$ & \parbox[c]{1.5cm}{$0.81_{-0.15}^{+0.33}$} & \parbox[c]{1.5cm}{$0.94_{-0.25}^{+0.39}$} \\
\vspace{5mm}
$S$ & \parbox[c]{1.5cm}{$0.019_{-0.013}^{+0.014}$} & \parbox[c]{1.5cm}{$0.013_{-0.009}^{+0.012}$} \\
\vspace{5mm}
$\log M_{\mathrm{HI}}$ & \parbox[c]{1.5cm}{$9.55_{-0.54}^{+0.35}$\\[1.5ex] $\left(9.47_{-0.32}^{+0.28}\right)\tnote{*}$} & \parbox[c]{1.5cm}{$9.72_{-0.73}^{+0.36}$\\[1.5ex] $\left(9.72_{-0.73}^{+0.36}\right)\tnote{*}$} \\
\vspace{5mm}
$\log (M_{\mathrm{HI}}/L_B$) & \parbox[c]{1.6cm}{$-0.48_{-0.25}^{+0.26}$\\[1.5ex]$\left(-0.57_{-0.12}^{+0.24}\right)\tnote{*}$} & \parbox[c]{1.6cm}{$-0.90_{-0.34}^{+0.35}$\\[1.5ex] $\left(-0.90_{-0.34}^{+0.35}\right)\tnote{*}$}\\
\vspace{5mm}
$C_i$ & \parbox[c]{1.5cm}{$2.12_{-0.15}^{+0.23}$} & \parbox[c]{1.5cm}{$2.52_{-0.24}^{+0.39}$} \\ 
\bottomrule
\end{tabular}
\begin{tablenotes}
\small
\item[*] The bracketed values for $M_{\mathrm{HI}}$ and $M_{\mathrm{HI}}/L_B$ represent the corresponding medians after missing values were imputed using the Iterative Imputer method, as discussed in Section \ref{sec:random forest}.
\end{tablenotes}
\end{threeparttable}
\label{tab:stat}
\end{table}
\end{nolinenumbers}

\begin{figure}[ht]
    \centering
    \begin{minipage}[t]{0.48\textwidth} 
        \centering
        \includegraphics[width=\linewidth]{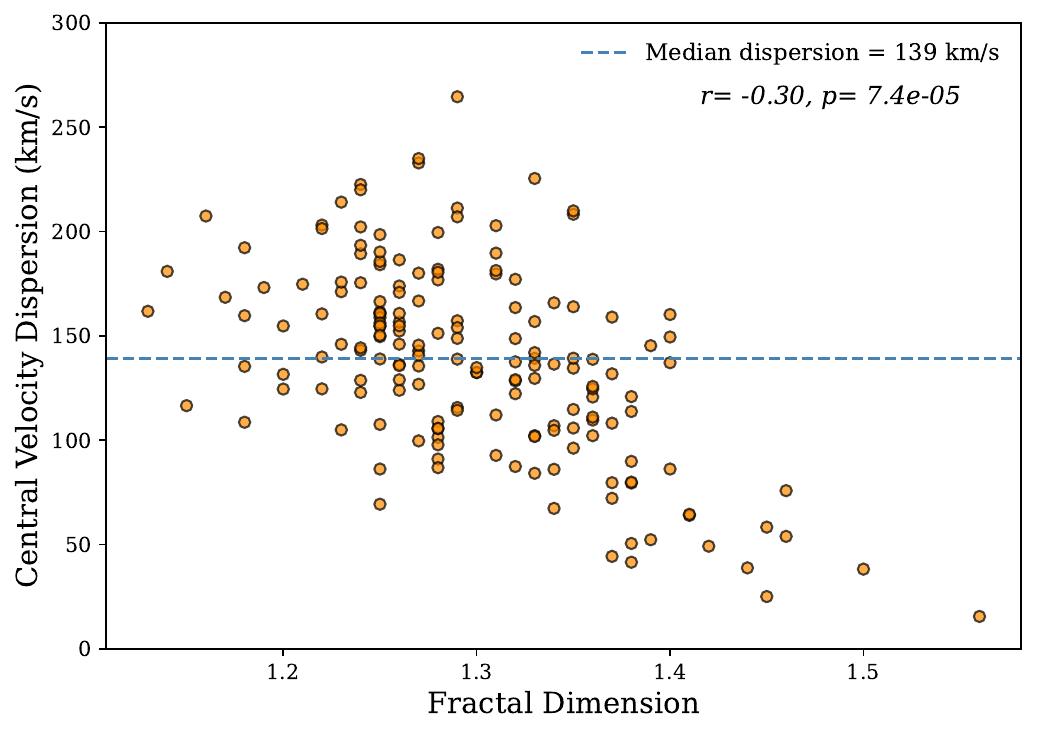}
    \end{minipage}
    \hfill  
    \begin{minipage}[t]{0.48\textwidth} 
        \centering
        \includegraphics[width=\linewidth]{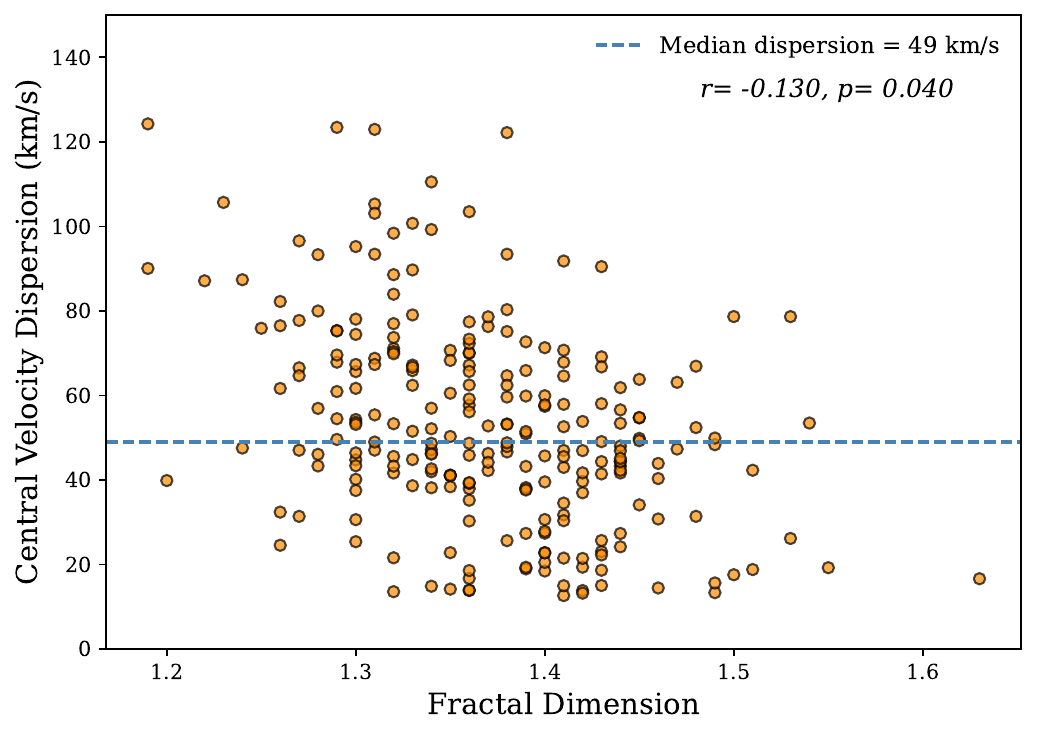}
    \end{minipage}
    \vspace{-6mm}
    \caption{Correlation between fractal dimension and central velocity dispersion (obtained from the SDSS \textit{specObj} table) for grand-designs (top) and flocculents (bottom).}
    \label{fig:correlation}
\end{figure}

\subsection{Physical interpretation}
We study the correlation of fractal dimension with the central velocity dispersion, which is measured from the spectra of the galaxies. The values for the central velocity dispersions are retrived from the SDSS \textit{specObj} table, which represents the velocity dispersion values in kms$^{-1}$ within the central $3''$ region of the galaxy. The presence of a bulge is indicated by high velocity dispersion in the central region of the galaxy, with a larger central velocity dispersion indicating the presence of a more massive bulge (\citealp{Ferrase..2000, Haring..2004}).\\

Figure \ref{fig:correlation} shows a negative correlation between fractal dimension and central velocity dispersion. While the correlation for the grand-designs is moderate and statistically significant ($r=-0.3$ , $p \ll 0.05$), the flocculents show a weaker correlation with low significance ($r=-0.12$ , $p \sim 0.05$). This is in agreement with their formation mechanisms. Grand-designs are believed to be driven by quasi-stationary density waves. The high Toomre Q parameter in the bulge provides the necessary Q-barrier that reflects incoming density waves before they reach the ILR. This allows the formation of a coherent, global two-armed spiral pattern characteristic of grand-design galaxies.  In contrast, galaxies
with weaker bulges have lower central Q values, which fail to reflect the waves strongly. As a result, the grand-design structure becomes less prominent. This explains the statistically significant negative correlation observed for the grand-designs. Flocculent spiral galaxies, on the other hand, are primarily driven by the local gravitational instabilities amplified by the swing amplification mechanism. This leads to their patchy, fragmented appearance. Unlike in the density wave theory, the presence of a bulge does not play any role in the development of flocculent spirals, which explains the insignificant correlation. This interpretation is also consistent with the observational findings by \citet{Bittner..2017}, who found grand-designs tend to host more massive bulges. We find that fractal dimension, is not only a powerful non-parametric measure defining the morphology of the spiral galaxies, but also a physically motivated quantity, connecting directly to the formation mechanism of the spiral features. 

\section{Conclusions}
\label{sec:conclusion}
We aim to introduce fractal dimension, calculated from the isophotes of spiral galaxies, as a robust metric to distinguish between flocculent and grand-design spirals. We use the i-band FITS images of 322 flocculent and 197 grand-design galaxies from SDSS DR18. The sample was obtained from the catalogs of \citet{Buta..2015} and \citet{Sarkar..2023}. We find the median values of the fractal dimension for grand-designs and flocculents to be $1.29_{-0.04}^{+0.06}$ and $1.38_{-0.06}^{+0.05}$, respectively. The fragmented appearance of flocculents is reflected in their relatively higher median value. Furthermore, a K-S test rejects the null hypothesis that the two distributions are drawn from the same population. \\

We then conduct a comparative study by incorporating several other parameters from the literature that have shown distinct distributions for the two spiral morphologies. One such parameter is the arm/inter-arm contrast $C$ introduced by \citet{Elmegreen..2011} to indicate the morphological-type  of the spiral arms. Grand-designs typically exhibit higher values of $C$ compared to flocculents. Another parameter included is the clumpiness $S$ parameter introduced by \citet{Conselice..2003}, which captures small-scale structures or clumps. We observe that flocculents, on average, have higher $S$ values than grand-designs, although the distinction between the two distributions is not very significant. These two parameters, like the fractal dimension, are non-parametric measures directly derived from CCD images of galaxies. In addition to these, we also include three other parameters: total atomic hydrogen mass ($M_{HI}$), ratio of atomic hydrogen mass-to-blue luminosity $M_{HI}/L_B$ and concentration index $C_i$.  We train a random forest model using our labeled sample and the six features mentioned above to evaluate their importance in classifying spiral morphologies. The fractal dimension outperforms all other parameters, achieving a feature importance of 30.8$\%$, followed by $C_i$ (26.0$\%$) and $M_{HI}$ (21.0$\%$). \\

Furthermore, we show that the fractal dimension for grand-designs exhibits a moderate and statistically significant correlation ( $r=-0.3$ , $p \ll 0.05$) with the central velocity dispersion, which serves as a proxy for bulge mass. This is in agreement with the density wave theory for grand-designs, which necessitates the presence of a massive bulge to reflect off the density waves and develop a prominent global two-armed spiral. In contrast, flocculent spirals show a very weak and statistically insignificant correlation ($r=-0.12$ , $p \sim 0.05$). This is consistent with the understanding that flocculents arise from local instabilities, where the presence of a bulge plays little to no role in their formation. Thus, the fractal dimension not only serves as a robust, non-parametric measure for classifying spiral morphologies but also captures the physical mechanisms underlying spiral structure formation.

\section*{Acknowledgments}
We thank the anonymous referee for their suggestions, which
helped to improve the clarity of the paper. We thank Prime Minister’s Research Fellowship (PMRF ID - 0903060) for funding this project.\\

Funding for the Sloan Digital Sky Survey V has been provided by the Alfred P. Sloan Foundation, the Heising-Simons Foundation, the National Science Foundation, and the Participating Institutions. SDSS acknowledges support and resources from the Center for High-Performance Computing at the University of Utah. SDSS telescopes are located at Apache Point Observatory, funded by the Astrophysical Research Consortium and operated by New Mexico State University, and at Las Campanas Observatory, operated by the Carnegie Institution for Science. The SDSS web site is \url{www.sdss.org}.\\

SDSS is managed by the Astrophysical Research Consortium for the Participating Institutions of the SDSS Collaboration, including Caltech, The Carnegie Institution for Science, Chilean National Time Allocation Committee (CNTAC) ratified researchers, The Flatiron Institute, the Gotham Participation Group, Harvard University, Heidelberg University, The Johns Hopkins University, L’Ecole polytechnique f\'{e}d\'{e}rale de Lausanne (EPFL), Leibniz-Institut f\"{u}r Astrophysik Potsdam (AIP), Max-Planck-Institut f\"{u}r Astronomie (MPIA Heidelberg), Max-Planck-Institut f\"{u}r Extraterrestrische Physik (MPE), Nanjing University, National Astronomical Observatories of China (NAOC), New Mexico State University, The Ohio State University, Pennsylvania State University, Smithsonian Astrophysical Observatory, Space Telescope Science Institute (STScI), the Stellar Astrophysics Participation Group, Universidad Nacional Aut\'{o}noma de M\'{e}xico, University of Arizona, University of Colorado Boulder, University of Illinois at Urbana-Champaign, University of Toronto, University of Utah, University of Virginia, Yale University, and Yunnan University.\\

\textit{Software}: \textsc{\texttt{astropy}} (\citealp{Astropy..2022}), \textsc{\texttt{matplotlib}} (\citealp{Matplotlib..2007}), \textsc{\texttt{numpy}} (\citealp{Numpy..2020}), and \textsc{\texttt{pandas}} (\citealp{pandas..2010}).

\bibliography{sample631}{}
\bibliographystyle{aasjournal}

\pagebreak

\appendix

\section{Values of Fractal dimension $D_B$ for the flocculent and grand-design samples.}
\label{appendix:fractal_values_appendix}

\begin{longtable}{p{1.6cm} p{0.7cm} p{0.4cm}@{\hspace{1.4cm}} p{1.6cm} p{0.7cm} p{0.4cm}@{\hspace{1.4cm}} p{1.6cm} p{0.7cm} p{0.4cm}}
\caption{Values of Fractal dimension $D_B$ for the samples of 156 flocculents and 58 grand-designs selected from \citet{Buta..2015}. `F' and `G' represent the flocculent and grand-design, respectively.}
\label{tab:fractal_values_buta}\\
\hline
\textbf{Name} & $\mathbf{D_B}$ & \textbf{Class} & \textbf{Name} & $\mathbf{D_B}$ & \textbf{Class} & \textbf{Name} & $\mathbf{D_B}$ & \textbf{Class}  \\
\hline

\endfirsthead

\hline
\textbf{Name} & $\mathbf{D_B}$ & \textbf{Class} & \textbf{Name} & $\mathbf{D_B}$ & \textbf{Class} & \textbf{Name} & $\mathbf{D_B}$ & \textbf{Class}  \\
\hline
\endhead

\hline
\multicolumn{9}{r}{{\textit{Table continues on the next page...}}} \\ 
\endfoot

\hline
\endlastfoot
ESO539-7 & 1.59 & F & ESO576-32 & 1.36 & F & IC1014 & 1.45 & F \\
IC1066 & 1.3 & F & IC1125 & 1.35 & F & IC1251 & 1.42 & F \\
IC797 & 1.48 & F & IC800 & 1.5 & F & NGC1051 & 1.47 & F \\
NGC1087 & 1.47 & F & NGC1299 & 1.3 & F & NGC2541 & 1.45 & F \\
NGC2552 & 1.63 & F & NGC2684 & 1.34 & F & NGC2701 & 1.34 & F \\
NGC2743 & 1.49 & F & NGC3020 & 1.44 & F & NGC3057 & 1.52 & F \\
NGC3153 & 1.4 & F & NGC3206 & 1.4 & F & NGC3225 & 1.3 & F \\
NGC3274 & 1.44 & F & NGC3299 & 1.55 & F & UGC7590 & 1.27 & F \\
UGC8041 & 1.45 & F & UGC8042 & 1.46 & F & UGC8053 & 1.55 & F \\
UGC8056 & 1.4 & F & UGC8516 & 1.36 & F & UGC8597 & 1.44 & F \\
UGC8733 & 1.48 & F & UGC8909 & 1.46 & F & UGC9215 & 1.38 & F \\
UGC9299 & 1.44 & F & UGC9936 & 1.2 & F & NGC3629 & 1.41 & F \\
NGC3659 & 1.41 & F & NGC3664 & 1.45 & F & NGC3755 & 1.41 & F \\
NGC3782 & 1.39 & F & NGC3794 & 1.42 & F & NGC3795A & 1.46 & F \\
NGC3846A & 1.49 & F & NGC3850 & 1.43 & F & NGC3876 & 1.28 & F \\
NGC3949 & 1.38 & F & NGC3985 & 1.34 & F & NGC4020 & 1.15 & F \\
NGC4032 & 1.36 & F & NGC4108B & 1.4 & F & NGC4141 & 1.36 & F \\
NGC4142 & 1.46 & F & NGC4288 & 1.42 & F & NGC428 & 1.44 & F \\
NGC4353 & 1.41 & F & NGC4376 & 1.47 & F & NGC4384 & 1.32 & F \\
NGC4385 & 1.31 & F & NGC4390 & 1.45 & F & NGC4409 & 1.43 & F \\
NGC4470 & 1.41 & F & NGC4502 & 1.4 & F & NGC450 & 1.43 & F \\
NGC4519 & 1.44 & F & NGC4525 & 1.49 & F & NGC4534 & 1.42 & F \\
NGC4595 & 1.38 & F & NGC4618 & 1.4 & F & NGC4630 & 1.35 & F \\
NGC4635 & 1.48 & F & NGC4658 & 1.4 & F & NGC4668 & 1.06 & F \\
NGC4765 & 1.26 & F & NGC4904 & 1.38 & F & NGC5117 & 1.42 & F \\
NGC5300 & 1.46 & F & NGC5334 & 1.55 & F & NGC5346 & 1.39 & F \\
NGC5486 & 1.39 & F & NGC5585 & 1.44 & F & NGC5624 & 1.43 & F \\
NGC5645 & 1.37 & F & NGC5667 & 1.4 & F & NGC5678 & 1.36 & F \\
NGC5713 & 1.41 & F & NGC5774 & 1.42 & F & NGC5798 & 1.38 & F \\
NGC5915 & 1.36 & F & NGC5949 & 1.46 & F & NGC3389 & 1.42 & F \\
NGC3455 & 1.43 & F & NGC3488 & 1.41 & F & NGC6155 & 1.36 & F \\
NGC6267 & 1.39 & F & NGC941 & 1.51 & F & PGC35705 & 1.49 & F \\
PGC43020 & 1.59 & F & UGC10290 & 1.51 & F & UGC10445 & 1.48 & F \\
UGC10854 & 1.42 & F & UGC12151 & 1.54 & F & UGC12682 & 1.51 & F \\
UGC1551 & 1.65 & F & UGC1862 & 1.52 & F & UGC2081 & 1.53 & F \\
UGC4543 & 1.5 & F & UGC5740 & 1.52 & F & UGC5832 & 1.43 & F \\
UGC5934 & 1.38 & F & UGC6162 & 1.43 & F & UGC6713 & 1.53 & F \\
UGC6900 & 1.5 & F & UGC7133 & 1.44 & F & IC2604 & 1.42 & F \\
NGC2537 & 1.42 & F & NGC3213 & 1.34 & F & NGC3346 & 1.5 & F \\
NGC3381 & 1.39 & F & NGC3445 & 1.35 & F & NGC3913 & 1.42 & F \\
NGC4037 & 1.44 & F & NGC4108 & 1.22 & F & NGC4136 & 1.34 & F \\
NGC4234 & 1.42 & F & NGC4276 & 1.44 & F & NGC4561 & 1.36 & F \\
NGC4571 & 1.38 & F & NGC4900 & 1.35 & F & NGC4961 & 1.3 & F \\
NGC5569 & 1.3 & F & NGC5600 & 1.32 & F & NGC5668 & 1.39 & F \\
NGC5789 & 1.51 & F & NGC5958 & 1.3 & F & NGC7625 & 1.33 & F \\
NGC991 & 1.41 & F & PGC42868 & 1.54 & F & PGC68771 & 1.44 & F \\
UGC10020 & 1.44 & F & UGC10437 & 1.36 & F & UGC10791 & 1.55 & F \\
UGC5172 & 1.44 & F & UGC5478 & 1.56 & F & UGC5707 & 1.38 & F \\
UGC5976 & 1.44 & F & UGC6249 & 1.43 & F & UGC6320 & 1.34 & F \\
UGC6849 & 1.38 & F & UGC6930 & 1.39 & F & UGC7690 & 1.42 & F \\
UGC8084 & 1.43 & F & UGC8153 & 1.44 & F & UGC8588 & 1.34 & F \\
UGC9569 & 1.35 & F & UGC9661 & 1.43 & F & UGC9875 & 1.49 & F \\
ESO576-1 & 1.26 & G & IC1151 & 1.38 & G & IC3102 & 1.33 & G \\
IC3115 & 1.45 & G & IC750 & 1.4 & G & IC769 & 1.5 & G \\
NGC1022 & 1.37 & G & NGC2543 & 1.38 & G & NGC2710 & 1.44 & G \\
NGC2780 & 1.42 & G & NGC2854 & 1.37 & G & NGC2856 & 1.36 & G \\
NGC2964 & 1.43 & G & NGC3049 & 1.38 & G & NGC3177 & 1.28 & G \\
NGC3185 & 1.32 & G & NGC3227 & 1.16 & G & NGC3583 & 1.38 & G \\
NGC3626 & 1.33 & G & NGC3689 & 1.35 & G & NGC3718 & 1.38 & G \\
NGC4165 & 1.38 & G & NGC4258 & 1.36 & G & NGC4413 & 1.39 & G \\
NGC4450 & 1.33 & G & NGC4531 & 1.45 & G & NGC4579 & 1.24 & G \\
NGC4795 & 1.37 & G & NGC5205 & 1.38 & G & NGC5248 & 1.28 & G \\
NGC5347 & 1.38 & G & NGC5448 & 1.32 & G & NGC5661 & 1.37 & G \\
NGC5850 & 1.35 & G & NGC5899 & 1.4 & G & NGC5950 & 1.41 & G \\
NGC6181 & 1.41 & G & NGC7479 & 1.36 & G & PGC11248 & 1.46 & G \\
UGC2443 & 1.47 & G & UGC4621 & 1.25 & G & UGC6309 & 1.46 & G \\
UGC8658 & 1.56 & G & NGC1068 & 1.19 & G & NGC2681 & 1.04 & G \\
NGC3433 & 1.41 & G & NGC3504 & 1.32 & G & NGC3507 & 1.34 & G \\
NGC3982 & 1.34 & G & NGC4314 & 1.3 & G & NGC4378 & 1.3 & G \\
NGC4412 & 1.37 & G & NGC5339 & 1.34 & G & NGC5957 & 1.25 & G \\
NGC718 & 1.3 & G & NGC7743 & 1.28 & G & NGC7798 & 1.28 & G \\
UGC6903 & 1.46 & G &  & &  &  & &  \\
\end{longtable}

\begin{longtable}{p{3.0cm} p{0.7cm} p{0.4cm}@{\hspace{1.4cm}} p{3.0cm} p{0.7cm} p{0.4cm}@{\hspace{1.4cm}} p{3.0cm} p{0.7cm} p{0.4cm}}
\caption{Sample as \ref{tab:fractal_values_buta}, but for the samples of \citet{Sarkar..2023}: 166 flocculents (F) and 139 grand-designs (G).}
\label{tab: fractal_values_suman}\\
\hline
\textbf{SDSS Name} & $\mathbf{D_B}$ & \textbf{Class} & \textbf{SDSS Name} & $\mathbf{D_B}$ & \textbf{Class} & \textbf{SDSS Name} & $\mathbf{D_B}$ & \textbf{Class}  \\
\hline

\endfirsthead

\hline
\textbf{Name} & $\mathbf{D_B}$ & \textbf{Class} & \textbf{Name} & $\mathbf{D_B}$ & \textbf{Class} & \textbf{Name} & $\mathbf{D_B}$ & \textbf{Class}  \\
\hline
\endhead

\hline
\multicolumn{9}{r}{{\textit{Table continues on the next page...}}} \\ 
\endfoot

\hline
\endlastfoot
J083322.68+523156.1 & 1.32 & F & J101942.81+572524.4 & 1.32 & F & J101511.42+564019.5 & 1.32 & F \\
J111316.56+590050.0 & 1.45 & F & J102016.67+243550.9 & 1.45 & F & J141635.20+095907.8 & 1.38 & F \\
J140121.81+102851.3 & 1.36 & F & J132956.71+110419.5 & 1.32 & F & J142502.97+274526.7 & 1.3 & F \\
J095409.24+582027.4 & 1.45 & F & J084455.20+474444.8 & 1.43 & F & J163849.53+172112.0 & 1.4 & F \\
J152136.81+111526.6 & 1.36 & F & J155100.98+514713.9 & 1.29 & F & J100513.48+212721.4 & 1.41 & F \\
J133457.26+340238.6 & 1.34 & F & J120825.58+100100.0 & 1.41 & F & J135507.96+401003.3 & 1.43 & F \\
J111629.13+410441.9 & 1.38 & F & J121821.43+251300.3 & 1.35 & F & J094549.96+282822.8 & 1.34 & F \\
J130837.55+540427.7 & 1.43 & F & J100351.02+502052.9 & 1.33 & F & J133231.22+044809.0 & 1.34 & F \\
J135231.94+374902.7 & 1.3 & F & J104527.33+583535.4 & 1.35 & F & J174215.28+555910.6 & 1.29 & F \\
J145934.75+325028.8 & 1.32 & F & J141503.69+362726.1 & 1.45 & F & J162122.05+404837.8 & 1.43 & F \\
J111814.71+263713.9 & 1.36 & F & J155220.73+243735.7 & 1.32 & F & J141516.22+342054.1 & 1.37 & F \\
J102625.53+173037.3 & 1.44 & F & J142938.52+103502.6 & 1.31 & F & J152356.52+380719.7 & 1.32 & F \\
J140520.32+304841.6 & 1.31 & F & J131211.30+212418.6 & 1.41 & F & J102929.32+193722.0 & 1.19 & F \\
J131432.44+304220.8 & 1.31 & F & J125137.96+312109.9 & 1.39 & F & J084132.11+511446.6 & 1.36 & F \\
J081759.80+463414.7 & 1.3 & F & J074618.81+390400.6 & 1.36 & F & J112945.19+220735.5 & 1.34 & F \\
J161145.91+372740.3 & 1.35 & F & J152352.72+233251.2 & 1.38 & F & J093249.24+622012.3 & 1.32 & F \\
J083834.00+304755.2 & 1.32 & F & J120520.97+630923.8 & 1.35 & F & J130213.92+670454.8 & 1.35 & F \\
J105039.84+654337.9 & 1.45 & F & J130910.84+222758.8 & 1.27 & F & J145449.24+180621.4 & 1.39 & F \\
J144252.58+183712.9 & 1.32 & F & J132140.53+312103.5 & 1.38 & F & J104554.26+371103.3 & 1.36 & F \\
J141452.03+140733.1 & 1.32 & F & J104351.09+212806.0 & 1.37 & F & J133455.34+312336.5 & 1.4 & F \\
J131258.27+311530.9 & 1.4 & F & J131730.95+310547.3 & 1.47 & F & J095106.02+090030.9 & 1.48 & F \\
J145756.35+534705.8 & 1.52 & F & J105300.09+173429.4 & 1.53 & F & J104509.23+185822.7 & 1.4 & F \\
J104622.05+131317.1 & 1.35 & F & J102138.25+123433.9 & 1.38 & F & J150429.73+021958.9 & 1.34 & F \\
J113926.83+032816.0 & 1.29 & F & J093832.18+371129.5 & 1.44 & F & J104559.59+224914.4 & 1.3 & F \\
J142612.19+483350.3 & 1.38 & F & J160216.34+491211.4 & 1.42 & F & J133026.37+300144.4 & 1.3 & F \\
J080627.40+505717.3 & 1.37 & F & J090758.29+414232.1 & 1.34 & F & J091639.79+071558.9 & 1.4 & F \\
J134459.54+292535.8 & 1.32 & F & J114327.28+524240.0 & 1.33 & F & J102810.12+495837.2 & 1.33 & F \\
J153433.38+410755.8 & 1.38 & F & J112104.80+311508.1 & 1.41 & F & J110658.99+230022.7 & 1.36 & F \\
J105828.10+242231.5 & 1.41 & F & J132113.05+311318.6 & 1.44 & F & J160149.70+080853.8 & 1.4 & F \\
J144842.57+122725.9 & 1.29 & F & J172408.08+585942.2 & 1.44 & F & J144349.20+532402.3 & 1.4 & F \\
J133248.69+415218.5 & 1.31 & F & J121746.92+105040.6 & 1.29 & F & J151542.84+012720.7 & 1.43 & F \\
J142709.50+305653.5 & 1.35 & F & J100105.40+165619.8 & 1.39 & F & J125822.89+451622.0 & 1.41 & F \\
J122553.80+451653.3 & 1.39 & F & J101027.91+275721.9 & 1.37 & F & J092032.71+174207.8 & 1.33 & F \\
J105045.14+465032.4 & 1.3 & F & J082049.32+223927.9 & 1.3 & F & J091301.27+202154.9 & 1.31 & F \\
J083054.11+201453.0 & 1.33 & F & J151808.40+051838.3 & 1.26 & F & J095900.21+174901.6 & 1.26 & F \\
J132327.45+062333.2 & 1.25 & F & J074438.73+402158.8 & 1.36 & F & J161918.14+370542.7 & 1.36 & F \\
J111856.18+001033.8 & 1.39 & F & J102132.73+223246.8 & 1.34 & F & J164208.18+401636.8 & 1.36 & F \\
J112924.07+345215.9 & 1.4 & F & J133037.05+411015.7 & 1.3 & F & J155218.55+232035.9 & 1.24 & F \\
J150709.50+093808.0 & 1.23 & F & J140814.40+354412.5 & 1.5 & F & J163103.88+410921.7 & 1.31 & F \\
J113050.49+603008.0 & 1.36 & F & J102524.21+553129.7 & 1.29 & F & J154122.58+281347.1 & 1.31 & F \\
J095011.30+161711.9 & 1.31 & F & J120050.49+315242.3 & 1.33 & F & J141311.61+130013.2 & 1.41 & F \\
J113245.43+405033.2 & 1.28 & F & J140830.45+362943.3 & 1.36 & F & J170443.10+343330.6 & 1.37 & F \\
J113244.06+614936.8 & 1.46 & F & J093225.05+572858.4 & 1.22 & F & J102829.39+535159.3 & 1.31 & F \\
J104038.59+371959.5 & 1.3 & F & J133602.56+661809.1 & 1.3 & F & J095524.67+410916.2 & 1.33 & F \\
J125945.33+320242.0 & 1.26 & F & J135814.18+363900.5 & 1.44 & F & J123807.68+224154.8 & 1.36 & F \\
J104839.19+214420.5 & 1.29 & F & J094423.49+111352.6 & 1.33 & F & J152602.66+161923.6 & 1.3 & F \\
J121505.37+133541.0 & 1.24 & F & J152145.69+230909.4 & 1.39 & F & J140436.76+291159.8 & 1.32 & F \\
J084015.50+560254.6 & 1.31 & F & J130727.96+522418.0 & 1.29 & F & J151024.21+450958.1 & 1.36 & F \\
J092455.63+555348.1 & 1.36 & F & J104615.88+510442.7 & 1.19 & F & J081618.48+522518.9 & 1.28 & F \\
J154006.76+204050.1 & 1.39 & F & J153840.43+410018.8 & 1.28 & F & J132341.56+313846.7 & 1.27 & F \\
J111224.71+312244.3 & 1.27 & F & J123848.02+320529.3 & 1.31 & F & J121058.75+431457.8 & 1.26 & F \\
J105659.30+181330.2 & 1.33 & F & J155555.36+170949.4 & 1.34 & F & J083203.50+240039.2 & 1.41 & F \\
J090720.92+402706.0 & 1.33 & F & J121227.95+390639.3 & 1.28 & F & J121008.52+390308.6 & 1.43 & F \\
J095410.67+021713.8 & 1.35 & F & J090007.89+165526.4 & 1.26 & F & J081915.52+244733.5 & 1.3 & F \\
J092101.68+390923.6 & 1.27 & F & J102841.90+034039.8 & 1.27 & F & J115524.71+391324.3 & 1.36 & F \\
J090933.47+323022.8 & 1.4 & F & J090617.35+500521.4 & 1.28 & G & J114404.89+600711.2 & 1.36 & G \\
J153350.64+051625.3 & 1.28 & G & J095557.04+133315.1 & 1.25 & G & J094453.63+225306.3 & 1.29 & G \\
J110606.88+295558.0 & 1.34 & G & J154645.21+310040.4 & 1.28 & G & J150501.03+054748.3 & 1.36 & G \\
J090125.79+070149.7 & 1.38 & G & J104238.11+235706.8 & 1.31 & G & J130612.74+252033.1 & 1.27 & G \\
J090131.19+491931.4 & 1.36 & G & J080642.79+390524.7 & 1.39 & G & J161126.44+292110.8 & 1.23 & G \\
J111628.05+291936.1 & 1.36 & G & J082046.07+163844.8 & 1.27 & G & J151747.35+071305.8 & 1.25 & G \\
J115844.63+281722.4 & 1.24 & G & J101556.01+485737.0 & 1.29 & G & J104710.75+025949.2 & 1.32 & G \\
J123106.76+522451.5 & 1.32 & G & J091335.59+122626.9 & 1.24 & G & J092609.43+491836.7 & 1.26 & G \\
J090719.58+461321.2 & 1.22 & G & J105341.80+384612.8 & 1.25 & G & J141848.50+105037.7 & 1.26 & G \\
J163456.87+254133.3 & 1.32 & G & J093710.07+165837.9 & 1.35 & G & J135603.40+133021.0 & 1.31 & G \\
J143722.15+363404.2 & 1.34 & G & J161854.59+350914.2 & 1.29 & G & J153642.16+433221.5 & 1.33 & G \\
J161953.49+513247.4 & 1.23 & G & J101725.41+644259.4 & 1.25 & G & J104152.94+211509.1 & 1.26 & G \\
J084009.47+522721.5 & 1.35 & G & J163743.75+251514.9 & 1.32 & G & J091601.77+173523.3 & 1.33 & G \\
J091346.29+621954.4 & 1.3 & G & J144911.08+110654.8 & 1.28 & G & J164022.68+334047.0 & 1.26 & G \\
J113606.64+621456.9 & 1.33 & G & J112911.08+423133.9 & 1.32 & G & J141623.85+393007.8 & 1.28 & G \\
J164641.59+360525.4 & 1.24 & G & J130117.11+535138.9 & 1.25 & G & J110441.90+041750.4 & 1.27 & G \\
J151858.17+204855.3 & 1.29 & G & J154630.00+320702.1 & 1.25 & G & J094357.45+414114.3 & 1.26 & G \\
J122246.75+655037.6 & 1.31 & G & J110602.35+042545.7 & 1.37 & G & J083517.47+554943.8 & 1.31 & G \\
J141823.95+262309.1 & 1.33 & G & J151018.40+074219.6 & 1.22 & G & J140932.37+144954.5 & 1.28 & G \\
J161403.28+141655.5 & 1.35 & G & J084922.55+364237.2 & 1.32 & G & J110338.42+451047.8 & 1.27 & G \\
J133036.95+345502.6 & 1.28 & G & J152552.58+074910.6 & 1.4 & G & J135531.29+264749.8 & 1.3 & G \\
J133117.44+292205.4 & 1.31 & G & J112938.61+353052.7 & 1.29 & G & J114223.73+101550.9 & 1.25 & G \\
J130411.85+611140.9 & 1.27 & G & J160151.95+154732.1 & 1.29 & G & J081321.11+575108.0 & 1.37 & G \\
J125524.83+521603.7 & 1.4 & G & J161026.54+121820.4 & 1.31 & G & J120514.73+381408.1 & 1.33 & G \\
J092528.24+233630.9 & 1.26 & G & J103228.17+271014.2 & 1.24 & G & J091937.94+272728.0 & 1.23 & G \\
J091933.11+055257.9 & 1.35 & G & J152854.93+074443.0 & 1.26 & G & J091503.52+291611.9 & 1.3 & G \\
J105907.08+050022.4 & 1.18 & G & J135805.66+214750.9 & 1.26 & G & J135222.75+213221.6 & 1.28 & G \\
J134215.02+015126.8 & 1.26 & G & J133823.47+065315.5 & 1.33 & G & J113704.17+240546.5 & 1.26 & G \\
J140217.59+074102.9 & 1.28 & G & J135232.73+075127.6 & 1.26 & G & J162412.19+300944.1 & 1.34 & G \\
J160619.53+162553.1 & 1.25 & G & J143838.28+523452.1 & 1.27 & G & J081422.03+391504.8 & 1.35 & G \\
J120428.94+300530.8 & 1.24 & G & J110032.51+020657.8 & 1.4 & G & J091046.43+332239.0 & 1.24 & G \\
J075719.70+111221.8 & 1.33 & G & J144429.23+044441.0 & 1.35 & G & J142714.56+112020.6 & 1.34 & G \\
J121422.05+560041.1 & 1.27 & G & J163134.53+403356.1 & 1.23 & G & J142246.68+375942.8 & 1.29 & G \\
J131156.35+452614.7 & 1.27 & G & J105533.12+421759.8 & 1.25 & G & J082512.07+202005.0 & 1.28 & G \\
J100547.32+142019.4 & 1.19 & G & J114517.56+264602.6 & 1.25 & G & J092040.65+150603.4 & 1.29 & G \\
J145459.83+621633.3 & 1.15 & G & J093011.76+555108.7 & 1.22 & G & J114425.77+332118.2 & 1.25 & G \\
J154934.29+134942.8 & 1.2 & G & J150411.78+072723.1 & 1.16 & G & J105259.06+373648.2 & 1.18 & G \\
J100447.30+573605.8 & 1.26 & G & J160915.71+254244.8 & 1.25 & G & J131432.68+443027.0 & 1.22 & G \\
J091504.79+415948.9 & 1.36 & G & J105310.05+372528.7 & 1.34 & G & J093637.36+342847.2 & 1.33 & G \\
J112306.93+651506.6 & 1.2 & G & J143348.33+035724.7 & 1.18 & G & J155946.58+370214.2 & 1.13 & G \\
J160142.96+484940.5 & 1.14 & G & J090958.07+621450.4 & 1.35 & G & J081314.20+522731.4 & 1.24 & G \\
J110047.95+104341.3 & 1.22 & G & J080328.94+332744.6 & 1.18 & G & J090022.63+645451.7 & 1.25 & G \\
J161711.61+495751.9 & 1.24 & G & J162415.16+201100.7 & 1.24 & G & J141057.23+252950.0 & 1.21 & G \\
J163038.87+350323.0 & 1.24 & G & J094508.97+683540.4 & 1.25 & G & J084141.20+403926.8 & 1.17 & G \\
J125813.56+262536.7 & 1.2 & G & J133130.19+333223.4 & 1.28 & G & J105100.31+121714.5 & 1.29 & G \\
J113914.88+170837.1 & 1.25 & G & J090649.87+484620.9 & 1.25 & G & J155801.84+145748.8 & 1.27 & G \\
J154812.64+111646.2 & 1.27 & G & J115122.68+532639.0 & 1.23 & G &  & &  \\

\end{longtable}

\section{Parameter values for the flocculent and grand-design samples used for the random forest}
\label{appendix:random_forest_appendix}

The values for the six parameters $D_B$, $C$, $S$, $M_{HI}$, $M_{HI}/L_B$, $C_i$ for the samples of flocculent and grand-design spirals used in the random forest analysis. As described in text, we restrict to galaxies with inclination angle $i \leq $ 50 \textdegree for the random forest analysis. Table \ref{tab:parameter_value_random} contains the information computed for the samples of 182 flocculents and 97 grand-designs, which has $i \leq$ 50 \textdegree. Some samples contain missing values of $\mathrm{log}(M_{HI})$ and $\mathrm{log}(M_{HI}/L_B)$ which could not be computed because of missing HI information on Hyperleda.

\begin{longtable}{cccccccc}
\caption{Values of the six parameters $D_B$, $C$, $S$, $M_{HI}$, $M_{HI}/L_B$, $C_i$ for the samples of 182 flocculents and 97 grand-designs selected on the basis of inclination angle $i \leq 50 \textdegree$. The last column represents the class: `F' for flocculent and `G' for grand-design.}\\
\hline
\textbf{SDSS Name} & $\mathbf{D_B}$ & $\mathbf{C}$ & $\mathbf{S}$ &
$\mathbf{\mathrm{log}(M_{HI})}$ & $\mathbf{\mathrm{log}(M_{HI}/L_B)}$& $\mathbf{C_i}$ & \textbf{class}\\
\noalign{\vskip 8pt}
\hline
\noalign{\vskip 3pt}
\endfirsthead

\hline
\noalign{\vskip 8pt}
\textbf{Name} & $\mathbf{D_B}$ & $\mathbf{C}$ & $\mathbf{S}$  &
$\mathbf{\mathrm{log}(M_{HI})}$ & $\mathbf{\mathrm{log}(M_{HI}/L_B)}$ & $\mathbf{C_i}$ & \textbf{class}\\
\noalign{\vskip 8pt}
\hline
\noalign{\vskip 3pt}
\endhead

\hline
\multicolumn{8}{r}{{\textit{Table continues on the next page...}}} \\ 
\endfoot

\hline
\endlastfoot
ESO576-32           & 1.36                 & 0.73             & -0.015   & 8.34          & -1.19                              & 1.7  & F     \\
IC797               & 1.48                 & 0.71             & 0.017    & 8.78          & -1.03                              & 2.56 & F     \\
IC800               & 1.5                  & 0.97             & -0.001   & 8.04          & -1.39                              & 2.0  & F     \\
NGC2684             & 1.34                 & 0.36             & 0.017    & 8.92          & -1.25                              & 2.06 & F     \\
NGC2701             & 1.34                 & 0.82             & 0.034    & 9.69          & -0.65                              & 1.93 & F     \\
NGC3299             & 1.55                 & 0.39             & -0.006   & 8.0           & -0.74                              & 1.73 & F     \\
UGC8516             & 1.36                 & 0.48             & 0.012    & 8.46          & -0.92                              & 2.06 & F     \\
UGC8909             & 1.46                 & 0.35             & -0.002   & 9.01          & -0.37                              & 2.03 & F     \\
NGC3782             & 1.39                 & 1.28             & 0.038    & 9.02          & -0.41                              & 2.09 & F     \\
NGC3795A            & 1.46                 & 0.78             & -0.006   & 9.08          & -0.2                               & 2.06 & F     \\
NGC3846A            & 1.49                 & 1.75             & -0.135   & 8.94          & -0.51                              & 2.47 & F     \\
NGC4032             & 1.36                 & 0.52             & 0.003    & -    & -                         & 2.33 & F     \\
NGC4108B            & 1.4                  & 1.05             & 0.011    & 9.65          & -0.16                              & 2.28 & F     \\
NGC4288             & 1.42                 & 1.54             & -0.018   & 9.59          & -0.2                               & 2.0  & F     \\
NGC428              & 1.44                 & 1.09             & 0.073    & 9.45          & -0.49                              & 2.02 & F     \\
NGC4353             & 1.41                 & 0.74             & 0.014    & 8.46          & -1.11                              & 1.88 & F     \\
NGC4384             & 1.32                 & 0.66             & 0.012    & 8.76          & -0.98                              & 2.61 & F     \\
NGC4390             & 1.45                 & 0.58             & 0.026    & 9.03          & -0.81                              & 2.18 & F     \\
NGC450              & 1.43                 & 1.82             & 0.047    & 9.19          & -0.65                              & 2.23 & F     \\
NGC4519             & 1.44                 & 1.58             & 0.055    & 9.62          & -0.34                              & 2.22 & F     \\
NGC4630             & 1.35                 & 1.2              & -0.01    & 8.66          & -0.91                              & 2.12 & F     \\
NGC4635             & 1.48                 & 0.73             & 0.022    & 8.56          & -0.93                              & 2.0  & F     \\
NGC4765             & 1.26                 & 0.65             & 0.016    & 8.84          & -0.34                              & 2.65 & F     \\
NGC5334             & 1.55                 & 1.71             & 0.096    & 9.37          & -0.51                              & 2.1  & F     \\
NGC5713             & 1.41                 & 1.87             & 0.044    & 9.89          & -0.82                              & 2.51 & F     \\
NGC5915             & 1.36                 & 1.93             & 0.048    & 9.8           & -0.77                              & 2.55 & F     \\
NGC6267             & 1.39                 & 0.81             & 0.024    & 9.44          & -0.77                              & 1.89 & F     \\
NGC941              & 1.51                 & 0.88             & 0.038    & 9.09          & -0.78                              & 2.34 & F     \\
UGC10445            & 1.48                 & 1.37             & 0.033    & 9.18          & -0.2                               & 2.13 & F     \\
UGC12682            & 1.51                 & 1.92             & 0.033    & 9.04          & -0.49                              & 2.54 & F     \\
UGC1551             & 1.65                 & 2.13             & 0.069    & 9.63          & -0.52                              & 2.48 & F     \\
UGC1862             & 1.52                 & 0.68             & 0.021    & 8.49          & -1.05                              & 1.94 & F     \\
UGC4543             & 1.5                  & 1.26             & 0.054    & 9.66          & 0.35                               & 1.99 & F     \\
UGC5832             & 1.43                 & 0.99             & 0.016    & 8.71          & -0.69                              & 2.14 & F     \\
UGC6713             & 1.53                 & 0.54             & -0.028   & 8.86          & 0.1                                & 1.71 & F     \\
IC2604              & 1.42                 & 1.29             & 0.017    & -    & -                         & 2.62 & F     \\
NGC2537             & 1.42                 & 1.39             & 0.043    & 8.24          & -0.87                              & 2.09 & F     \\
NGC3213             & 1.34                 & 0.39             & 0.007    & 8.21          & -1.08                              & 1.97 & F     \\
NGC3346             & 1.5                  & 0.98             & 0.062    & 9.07          & -0.79                              & 1.76 & F     \\
NGC3381             & 1.39                 & 1.31             & 0.025    & 9.24          & -0.72                              & 2.13 & F     \\
NGC3445             & 1.35                 & 0.81             & 0.026    & 9.62          & -0.52                              & 2.06 & F     \\
NGC3913             & 1.42                 & 0.87             & 0.006    & 8.89          & -0.57                              & 2.62 & F     \\
NGC4037             & 1.44                 & 1.68             & -0.072   & 8.38          & -1.05                              & 2.26 & F     \\
NGC4108             & 1.22                 & 0.46             & 0.015    & 9.65          & -0.69                              & 2.38 & F     \\
NGC4136             & 1.34                 & 0.73             & 0.005    & 9.03          & -0.61                              & 1.92 & F     \\
NGC4234             & 1.42                 & 0.99             & 0.023    & 8.92          & -0.97                              & 2.17 & F     \\
NGC4276             & 1.44                 & 0.74             & 0.136    & 8.96          & -0.86                              & 2.28 & F     \\
NGC4561             & 1.36                 & 0.83             & 0.021    & 9.42          & -0.44                              & 2.0  & F     \\
NGC4571             & 1.38                 & 1.17             & 0.035    & 8.94          & -1.2                               & 1.98 & F     \\
NGC4900             & 1.35                 & 0.79             & 0.108    & 8.92          & -0.99                              & 1.71 & F     \\
NGC4961             & 1.3                  & 0.92             & 0.026    & 9.68          & -0.22                              & 2.69 & F     \\
NGC5569             & 1.3                  & 0.47             & -0.357   & -    & -                         & 1.86 & F     \\
NGC5600             & 1.32                 & 0.64             & 0.029    & 9.3           & -0.94                              & 2.38 & F     \\
NGC5668             & 1.39                 & 1.33             & 0.049    & 9.68          & -0.57                              & 2.4  & F     \\
NGC5789             & 1.51                 & 1.17             & 0.032    & -    & -                         & 1.8  & F     \\
NGC5958             & 1.3                  & 0.77             & 0.03     & 9.04          & -0.99                              & 2.17 & F     \\
NGC7625             & 1.33                 & 0.77             & 0.026    & 9.29          & -0.63                              & 2.5  & F     \\
NGC991              & 1.41                 & 0.91             & -0.017   & 9.17          & -0.57                              & 1.97 & F     \\
PGC42868            & 1.54                 & 1.8              & 0.012    & 9.44          & -0.31                              & 1.92 & F     \\
PGC68771            & 1.44                 & 1.57             & 0.005    & 9.45          & 0.13                               & 2.1  & F     \\
UGC10020            & 1.44                 & 0.78             & -0.067   & 9.3           & -0.33                              & 2.26 & F     \\
UGC10437            & 1.36                 & 1.1              & 0.085    & 9.72          & 0.18                               & 2.68 & F     \\
UGC10791            & 1.55                 & 2.11             & 0.02     & 8.97          & -                         & 1.95 & F     \\
UGC5172             & 1.44                 & 1.73             & 0.052    & 9.29          & 0.02                               & 2.07 & F     \\
UGC5478             & 1.56                 & 1.48             & 0.208    & 8.82          & -0.25                              & 1.73 & F     \\
UGC5707             & 1.38                 & 1.15             & 0.088    & 9.29          & -0.07                              & 2.22 & F     \\
UGC5976             & 1.44                 & 1.04             & 0.002    & -    & -                         & 2.26 & F     \\
UGC6249             & 1.43                 & 0.7              & 0.008    & 9.0           & 0.01                               & 1.88 & F     \\
UGC6320             & 1.34                 & 0.61             & 0.01     & 8.5           & -0.81                              & 2.3  & F     \\
UGC6849             & 1.38                 & 0.41             & -0.021   & 8.54          & -0.11                              & 1.53 & F     \\
UGC6930             & 1.39                 & 0.79             & -0.049   & 9.14          & -0.44                              & 2.42 & F     \\
UGC7690             & 1.42                 & 1.05             & 0.018    & 8.86          & -0.33                              & 2.29 & F     \\
UGC8084             & 1.43                 & 1.79             & -0.083   & 9.2           & -0.36                              & 2.42 & F     \\
UGC8153             & 1.44                 & 1.34             & 0.135    & 9.51          & -0.25                              & 2.61 & F     \\
UGC8588             & 1.34                 & 0.63             & 0.067    & 8.78          & -0.16                              & 1.67 & F     \\
UGC9569             & 1.35                 & 0.87             & 0.033    & 9.55          & -0.21                              & 1.91 & F     \\
UGC9661             & 1.43                 & 0.78             & 0.021    & 8.27          & -0.76                              & 2.2  & F     \\
UGC9875             & 1.49                 & 1.4              & 0.033    & 9.06          & 0.25                               & 2.0  & F     \\
IC3115              & 1.45                 & 2.29             & 0.056    & 8.94          & -0.75                              & 2.25 & G     \\
IC769               & 1.5                  & 1.13             & 0.0      & 9.66          & -                         & 2.1  & G     \\
NGC2780             & 1.42                 & 0.41             & 0.015    & 8.41          & -1.21                              & 1.93 & G     \\
NGC2964             & 1.43                 & 0.94             & 0.028    & 9.23          & -0.93                              & 2.24 & G     \\
NGC3177             & 1.28                 & 0.64             & -0.031   & 8.58          & -1.17                              & 2.9  & G     \\
NGC4413             & 1.39                 & 0.63             & 0.004    & 8.23          & -1.37                              & 2.12 & G     \\
NGC4450             & 1.33                 & 0.35             & 0.007    & 8.54          & -2.0                               & 2.42 & G     \\
NGC4579             & 1.24                 & 0.32             & 0.016    & 8.99          & -1.82                              & 2.48 & G     \\
NGC5347             & 1.38                 & 0.84             & 0.011    & 9.67          & -0.46                              & 2.29 & G     \\
NGC5850             & 1.35                 & 0.43             & 0.006    & 9.77          & -1.09                              & 2.83 & G     \\
NGC7479             & 1.36                 & 3.44             & 0.033    & 9.88          & -0.97                              & 2.59 & G     \\
UGC6309             & 1.46                 & 1.04             & 0.025    & 9.2           & -0.89                              & 2.09 & G     \\
NGC1068             & 1.19                 & 0.86             & 0.036    & 8.82          & -1.67                              & 2.27 & G     \\
NGC2681             & 1.04                 & 0.95             & 0.121    & 7.44          & -2.82                              & 2.61 & G     \\
NGC3433             & 1.41                 & 2.03             & -0.018   & 9.84          & -                         & 1.91 & G     \\
NGC3504             & 1.32                 & 2.93             & 0.012    & 8.85          & -1.57                              & 2.97 & G     \\
NGC3507             & 1.34                 & 1.33             & 0.034    & 8.97          & -0.9                               & 2.39 & G     \\
NGC3982             & 1.34                 & 0.48             & 0.035    & 9.34          & -0.97                              & 2.31 & G     \\
NGC4314             & 1.3                  & 4.68             & 0.033    & -    & -                         & 3.17 & G     \\
NGC4378             & 1.3                  & 0.94             & -0.017   & 9.8           & -0.94                              & 3.51 & G     \\
NGC4412             & 1.37                 & 0.91             & 0.04     & 9.02          & -1.14                              & 1.96 & G     \\
NGC5339             & 1.34                 & 1.07             & 0.053    & 9.36          & -1.07                              & 2.03 & G     \\
NGC5957             & 1.25                 & 0.81             & 0.011    & 9.59          & -0.25                              & 1.97 & G     \\
NGC718              & 1.3                  & 1.18             & 0.014    & 7.43          & -2.64                              & 3.15 & G     \\
NGC7743             & 1.28                 & 1.05             & -0.001   & 8.43          & -1.51                              & 2.52 & G     \\
NGC7798             & 1.28                 & 0.78             & 0.031    & 9.05          & -1.27                              & 2.22 & G     \\
UGC6903             & 1.46                 & 1.64             & -0.958   & 9.45          & -0.37                              & 1.8  & G     \\
J101511.42+564019.5 & 1.32                 & 0.46             & 0.036    & -    & -                         & 2.25 & F     \\
J163849.53+172112.0 & 1.4                  & 0.55             & 0.018    & 9.34          & -0.65                              & 1.69 & F     \\
J100513.48+212721.4 & 1.41                 & 0.96             & 0.034    & 10.0          & -0.35                              & 2.2  & F     \\
J120825.58+100100.0 & 1.41                 & 0.73             & 0.022    & 9.71          & -0.48                              & 2.28 & F     \\
J111629.13+410441.9 & 1.38                 & 0.95             & 0.019    & -    & -                         & 2.35 & F     \\
J130837.55+540427.7 & 1.43                 & 2.21             & 0.016    & -    & -                         & 2.44 & F     \\
J174215.28+555910.6 & 1.29                 & 0.38             & -0.04    & -    & -                         & 2.09 & F     \\
J145934.75+325028.8 & 1.32                 & 0.6              & -0.086   & 9.52          & 0.24                               & 1.92 & F     \\
J141503.69+362726.1 & 1.45                 & 0.87             & 0.003    & 8.9           & -0.03                              & 1.6  & F     \\
J162122.05+404837.8 & 1.43                 & 0.98             & 0.008    & -    & -                         & 2.21 & F     \\
J141516.22+342054.1 & 1.37                 & 0.55             & 0.038    & 9.82          & -0.67                              & 2.22 & F     \\
J102625.53+173037.3 & 1.44                 & 0.87             & 0.032    & 9.98          & -0.51                              & 2.41 & F     \\
J152356.52+380719.7 & 1.32                 & 0.45             & 0.014    & -    & -                         & 2.32 & F     \\
J140520.32+304841.6 & 1.31                 & 0.63             & -0.001   & 9.67          & -0.44                              & 1.81 & F     \\
J102929.32+193722.0 & 1.19                 & 0.67             & 0.003    & 10.01         & -0.66                              & 2.73 & F     \\
J131432.44+304220.8 & 1.31                 & 0.42             & 0.018    & 10.39         & -0.41                              & 2.82 & F     \\
J081759.80+463414.7 & 1.3                  & 0.96             & 0.036    & -    & -                         & 2.11 & F     \\
J112945.19+220735.5 & 1.34                 & 0.96             & 0.023    & 9.94          & -0.59                              & 2.12 & F     \\
J152352.72+233251.2 & 1.38                 & 0.48             & -0.014   & 10.0          & -0.42                              & 1.61 & F     \\
J093249.24+622012.3 & 1.32                 & 0.81             & 0.014    & -    & -                         & 2.62 & F     \\
J105039.84+654337.9 & 1.45                 & 0.7              & 0.012    & 9.64          & -0.88                              & 1.84 & F     \\
J104351.09+212806.0 & 1.37                 & 0.81             & 0.059    & -    & -                         & 1.74 & F     \\
J133455.34+312336.5 & 1.4                  & 1.09             & 0.003    & 10.25         & 0.14                               & 2.54 & F     \\
J131258.27+311530.9 & 1.4                  & 1.22             & -0.002   & 9.71          & -0.51                              & 1.91 & F     \\
J095106.02+090030.9 & 1.48                 & 0.72             & 0.011    & 10.04         & -0.42                              & 2.14 & F     \\
J145756.35+534705.8 & 1.52                 & 1.12             & 0.034    & -    & -                         & 2.21 & F     \\
J105300.09+173429.4 & 1.53                 & 1.3              & -0.01    & -    & -                         & 2.07 & F     \\
J104622.05+131317.1 & 1.35                 & 0.82             & 0.026    & 10.3          & -0.11                              & 2.25 & F     \\
J150429.73+021958.9 & 1.34                 & 1.36             & -0.012   & -    & -                         & 2.4  & F     \\
J113926.83+032816.0 & 1.29                 & 0.63             & 0.007    & 10.01         & -0.31                              & 2.83 & F     \\
J093832.18+371129.5 & 1.44                 & 1.4              & -0.105   & -    & -                         & 2.81 & F     \\
J104559.59+224914.4 & 1.3                  & 0.8              & 0.006    & 10.31         & -0.3                               & 2.39 & F     \\
J090758.29+414232.1 & 1.34                 & 1.48             & 0.033    & -    & -                         & 2.23 & F     \\
J134459.54+292535.8 & 1.32                 & 0.62             & 0.01     & 10.21         & -0.33                              & 2.16 & F     \\
J112104.80+311508.1 & 1.41                 & 1.0              & 0.009    & 10.06         & -0.4                               & 3.02 & F     \\
J105828.10+242231.5 & 1.41                 & 1.63             & 0.027    & -    & -                         & 2.49 & F     \\
J144842.57+122725.9 & 1.29                 & 0.61             & 0.017    & 9.29          & -0.33                              & 3.52 & F     \\
J172408.08+585942.2 & 1.44                 & 1.42             & 0.01     & 9.58          & -0.45                              & 2.59 & F     \\
J100105.40+165619.8 & 1.39                 & 0.78             & 0.011    & 9.98          & -0.4                               & 2.36 & F     \\
J095557.04+133315.1 & 1.25                 & 0.67             & 0.024    & -    & -                         & 2.7  & G     \\
J094453.63+225306.3 & 1.29                 & 1.25             & 0.014    & -    & -                         & 2.48 & G     \\
J154645.21+310040.4 & 1.28                 & 1.44             & 0.029    & 10.18         & -0.37                              & 2.97 & G     \\
J090125.79+070149.7 & 1.38                 & 1.92             & -0.016   & -    & -                         & 2.12 & G     \\
J090131.19+491931.4 & 1.36                 & 0.83             & 0.001    & -    & -                         & 3.28 & G     \\
J080642.79+390524.7 & 1.39                 & 1.34             & 0.044    & 10.54         & -0.29                              & 3.03 & G     \\
J115844.63+281722.4 & 1.24                 & 0.31             & 0.006    & 10.47         & -                         & 3.55 & G     \\
J091335.59+122626.9 & 1.24                 & 0.44             & 0.016    & 9.56          & -0.99                              & 2.69 & G     \\
J090719.58+461321.2 & 1.22                 & 1.43             & 0.03     & -    & -                         & 4.03 & G     \\
J161953.49+513247.4 & 1.23                 & 0.98             & 0.049    & -    & -                         & 2.74 & G     \\
J091601.77+173523.3 & 1.33                 & 0.89             & -0.002   & -    & -                         & 2.59 & G     \\
J144911.08+110654.8 & 1.28                 & 0.86             & 0.014    & 10.46         & -0.49                              & 2.42 & G     \\
J141623.85+393007.8 & 1.28                 & 0.76             & 0.006    & -    & -                         & 3.05 & G     \\
J130117.11+535138.9 & 1.25                 & 0.5              & 0.027    & -    & -                         & 2.14 & G     \\
J094357.45+414114.3 & 1.26                 & 0.76             & 0.016    & -    & -                         & 2.5  & G     \\
J122246.75+655037.6 & 1.31                 & 2.62             & 0.011    & 8.85          & -1.48                              & 2.73 & G     \\
J161403.28+141655.5 & 1.35                 & 1.01             & -0.006   & -    & -                         & 2.45 & G     \\
J152552.58+074910.6 & 1.4                  & 2.39             & 0.029    & 10.12         & -0.77                              & 2.33 & G     \\
J133117.44+292205.4 & 1.31                 & 1.11             & -0.009   & 10.69         & -0.17                              & 3.02 & G     \\
J112938.61+353052.7 & 1.29                 & 1.27             & 0.009    & 9.8           & -0.9                               & 2.76 & G     \\
J130411.85+611140.9 & 1.27                 & 1.16             & -0.182   & -    & -                         & 3.18 & G     \\
J160151.95+154732.1 & 1.29                 & 1.18             & -0.003   & -    & -                         & 2.22 & G     \\
J125524.83+521603.7 & 1.4                  & 0.66             & 0.028    & -    & -                         & 1.94 & G     \\
J120514.73+381408.1 & 1.33                 & 0.57             & 0.014    & -    & -                         & 2.31 & G     \\
J092528.24+233630.9 & 1.26                 & 0.82             & 0.004    & -    & -                         & 2.58 & G     \\
J152854.93+074443.0 & 1.26                 & 0.78             & 0.054    & 10.02         & -0.79                              & 2.52 & G     \\
J134215.02+015126.8 & 1.26                 & 0.52             & 0.016    & -    & -                         & 2.55 & G     \\
J113704.17+240546.5 & 1.26                 & 0.99             & 0.018    & 10.05         & -0.75                              & 2.5  & G     \\
J110032.51+020657.8 & 1.4                  & 1.17             & 0.017    & 10.62         & -0.29                              & 2.4  & G     \\
J075719.70+111221.8 & 1.33                 & 0.98             & 0.018    & 10.14         & -0.56                              & 2.26 & G     \\
J105533.12+421759.8 & 1.25                 & 0.44             & 0.0      & -    & -                         & 3.25 & G     \\
J091301.27+202154.9 & 1.31                 & 0.69             & 0.016    & 9.42          & 0.0                                & 1.68 & F     \\
J083054.11+201453.0 & 1.33                 & 0.62             & -0.002   & -    & -                         & 1.87 & F     \\
J151808.40+051838.3 & 1.26                 & 0.58             & 0.028    & 10.2          & -0.73                              & 2.17 & F     \\
J095900.21+174901.6 & 1.26                 & 0.8              & 0.023    & -    & -                         & 2.28 & F     \\
J132327.45+062333.2 & 1.25                 & 0.8              & 0.008    & 10.02         & -0.6                               & 1.98 & F     \\
J074438.73+402158.8 & 1.36                 & 0.79             & 0.005    & 9.79          & -0.08                              & 1.76 & F     \\
J161918.14+370542.7 & 1.36                 & 0.97             & 0.028    & -    & -                         & 2.31 & F     \\
J111856.18+001033.8 & 1.39                 & 1.3              & 0.023    & 10.13         & -0.23                              & 2.5  & F     \\
J102132.73+223246.8 & 1.34                 & 0.98             & -0.012   & 9.99          & -0.17                              & 2.23 & F     \\
J164208.18+401636.8 & 1.36                 & 0.77             & 0.014    & -    & -                         & 2.04 & F     \\
J112924.07+345215.9 & 1.4                  & 1.4              & 0.013    & -    & -                         & 1.97 & F     \\
J133037.05+411015.7 & 1.3                  & 0.86             & -0.018   & -    & -                         & 2.03 & F     \\
J155218.55+232035.9 & 1.24                 & 0.67             & 0.023    & 9.67          & -0.87                              & 1.98 & F     \\
J150709.50+093808.0 & 1.23                 & 0.7              & 0.015    & 10.48         & -0.37                              & 2.1  & F     \\
J140814.40+354412.5 & 1.5                  & 3.06             & 0.084    & 10.19         & 0.2                                & 1.96 & F     \\
J163103.88+410921.7 & 1.31                 & 0.7              & -0.008   & 10.1          & -0.65                              & 2.38 & F     \\
J113050.49+603008.0 & 1.36                 & 0.63             & 0.054    & 9.07          & -0.17                              & 1.91 & F     \\
J102524.21+553129.7 & 1.29                 & 0.57             & -0.003   & -    & -                         & 1.98 & F     \\
J154122.58+281347.1 & 1.31                 & 0.73             & 0.026    & -    & -                         & 2.07 & F     \\
J095011.30+161711.9 & 1.31                 & 0.68             & 0.021    & 9.82          & -0.4                               & 1.9  & F     \\
J120050.49+315242.3 & 1.33                 & 0.5              & 0.021    & 9.95          & -0.35                              & 1.78 & F     \\
J141311.61+130013.2 & 1.41                 & 0.59             & 0.019    & 9.95          & -0.3                               & 1.73 & F     \\
J113245.43+405033.2 & 1.28                 & 1.7              & -0.01    & -    & -                         & 2.54 & F     \\
J140830.45+362943.3 & 1.36                 & 0.88             & 0.029    & -    & -                         & 2.29 & F     \\
J170443.10+343330.6 & 1.37                 & 1.22             & 0.035    & -    & -                         & 2.1  & F     \\
J113244.06+614936.8 & 1.46                 & 0.54             & 0.006    & 9.24          & -0.81                              & 1.86 & F     \\
J093225.05+572858.4 & 1.22                 & 0.9              & 0.023    & -    & -                         & 2.07 & F     \\
J102829.39+535159.3 & 1.31                 & 0.96             & 0.023    & -    & -                         & 2.29 & F     \\
J104038.59+371959.5 & 1.3                  & 0.56             & 0.015    & -    & -                         & 1.82 & F     \\
J133602.56+661809.1 & 1.3                  & 1.17             & 0.076    & 9.6           & 0.05                               & 2.07 & F     \\
J095524.67+410916.2 & 1.33                 & 0.49             & 0.022    & -    & -                         & 2.07 & F     \\
J125945.33+320242.0 & 1.26                 & 0.75             & 0.015    & 9.76          & -0.69                              & 2.1  & F     \\
J135814.18+363900.5 & 1.44                 & 1.37             & -0.019   & 9.85          & 0.25                               & 2.61 & F     \\
J123807.68+224154.8 & 1.36                 & 0.82             & 0.025    & 9.56          & -0.57                              & 2.14 & F     \\
J104839.19+214420.5 & 1.29                 & 0.86             & 0.088    & 9.99          & -0.05                              & 2.23 & F     \\
J094423.49+111352.6 & 1.33                 & 0.67             & 0.091    & 9.72          & -0.29                              & 2.09 & F     \\
J152602.66+161923.6 & 1.3                  & 0.85             & 0.029    & 9.91          & -0.68                              & 1.97 & F     \\
J121505.37+133541.0 & 1.24                 & 0.86             & 0.017    & 9.82          & -0.83                              & 1.96 & F     \\
J152145.69+230909.4 & 1.39                 & 0.77             & 0.298    & 10.09         & -0.23                              & 1.81 & F     \\
J140436.76+291159.8 & 1.32                 & 0.87             & -0.959   & 9.86          & -0.45                              & 1.98 & F     \\
J084015.50+560254.6 & 1.31                 & 0.56             & 0.019    & -    & -                         & 1.97 & F     \\
J130727.96+522418.0 & 1.29                 & 0.79             & -0.001   & -    & -                         & 2.39 & F     \\
J151024.21+450958.1 & 1.36                 & 0.66             & 0.045    & -    & -                         & 2.05 & F     \\
J092455.63+555348.1 & 1.36                 & 0.61             & 0.015    & -    & -                         & 2.13 & F     \\
J104615.88+510442.7 & 1.19                 & 0.55             & 0.204    & -    & -                         & 3.32 & F     \\
J081618.48+522518.9 & 1.28                 & 0.64             & 0.005    & -    & -                         & 2.47 & F     \\
J154006.76+204050.1 & 1.39                 & 0.58             & 0.019    & 9.77          & -0.53                              & 1.93 & F     \\
J153840.43+410018.8 & 1.28                 & 0.82             & 0.006    & 9.91          & 0.05                               & 1.8  & F     \\
J132341.56+313846.7 & 1.27                 & 0.73             & 0.014    & 8.94          & -1.0                               & 2.27 & F     \\
J111224.71+312244.3 & 1.27                 & 0.6              & 0.019    & 10.1          & -0.61                              & 2.0  & F     \\
J123848.02+320529.3 & 1.31                 & 0.77             & 0.025    & 9.56          & -0.77                              & 2.15 & F     \\
J121058.75+431457.8 & 1.26                 & 0.68             & 0.009    & -    & -                         & 2.75 & F     \\
J105659.30+181330.2 & 1.33                 & 0.85             & 0.011    & 10.25         & -0.23                              & 2.09 & F     \\
J155555.36+170949.4 & 1.34                 & 1.18             & 0.019    & 10.12         & 0.0                                & 1.89 & F     \\
J083203.50+240039.2 & 1.41                 & 0.98             & 0.023    & 9.97          & -0.65                              & 2.16 & F     \\
J090720.92+402706.0 & 1.33                 & 0.66             & 0.069    & -    & -                         & 2.35 & F     \\
J121227.95+390639.3 & 1.28                 & 1.09             & 0.011    & -    & -                         & 2.81 & F     \\
J121008.52+390308.6 & 1.43                 & 0.78             & 0.023    & -    & -                         & 2.02 & F     \\
J095410.67+021713.8 & 1.35                 & 0.71             & 0.019    & 9.97          & -0.51                              & 2.02 & F     \\
J090007.89+165526.4 & 1.26                 & 0.96             & 0.044    & 9.97          & -0.06                              & 1.91 & F     \\
J081915.52+244733.5 & 1.3                  & 0.52             & 0.05     & 9.76          & -0.47                              & 2.17 & F     \\
J092101.68+390923.6 & 1.27                 & 0.5              & 0.055    & -    & -                         & 2.02 & F     \\
J102841.90+034039.8 & 1.27                 & 0.71             & 0.01     & 10.0          & -0.54                              & 2.33 & F     \\
J115524.71+391324.3 & 1.36                 & 0.53             & 0.056    & 9.82          & -0.17                              & 2.03 & F     \\
J090933.47+323022.8 & 1.4                  & 1.96             & 0.088    & -    & -                         & 2.87 & F     \\
J082512.07+202005.0 & 1.28                 & 0.72             & 0.021    & 8.95          & -1.4                               & 1.94 & G     \\
J100547.32+142019.4 & 1.19                 & 0.69             & 0.059    & -    & -                         & 2.42 & G     \\
J114517.56+264602.6 & 1.25                 & 1.47             & 0.02     & 10.01         & -0.66                              & 2.49 & G     \\
J092040.65+150603.4 & 1.29                 & 0.76             & 0.014    & 9.82          & -0.58                              & 2.63 & G     \\
J145459.83+621633.3 & 1.15                 & 0.46             & -0.003   & -    & -                         & 3.55 & G     \\
J093011.76+555108.7 & 1.22                 & 3.01             & 0.006    & -    & -                         & 4.48 & G     \\
J114425.77+332118.2 & 1.25                 & 0.85             & 0.009    & 9.97          & -0.73                              & 2.44 & G     \\
J154934.29+134942.8 & 1.2                  & 1.34             & 0.004    & -    & -                         & 2.89 & G     \\
J150411.78+072723.1 & 1.16                 & 0.44             & 0.006    & -    & -                         & 2.99 & G     \\
J105259.06+373648.2 & 1.18                 & 0.55             & 0.005    & -    & -                         & 2.63 & G     \\
J100447.30+573605.8 & 1.26                 & 2.11             & 0.026    & -    & -                         & 2.76 & G     \\
J160915.71+254244.8 & 1.25                 & 1.0              & 0.009    & -    & -                         & 2.19 & G     \\
J131432.68+443027.0 & 1.22                 & 3.38             & 0.012    & -    & -                         & 2.49 & G     \\
J091504.79+415948.9 & 1.36                 & 2.76             & 0.003    & -    & -                         & 3.04 & G     \\
J105310.05+372528.7 & 1.34                 & 1.13             & 0.008    & -    & -                         & 3.19 & G     \\
J093637.36+342847.2 & 1.33                 & 3.77             & 0.087    & -    & -                         & 2.72 & G     \\
J112306.93+651506.6 & 1.2                  & 0.48             & 0.014    & -    & -                         & 1.97 & G     \\
J143348.33+035724.7 & 1.18                 & 1.25             & 0.004    & -    & -                         & 2.73 & G     \\
J155946.58+370214.2 & 1.13                 & 0.56             & 0.012    & -    & -                         & 2.28 & G     \\
J160142.96+484940.5 & 1.14                 & 0.66             & 0.01     & -    & -                         & 2.91 & G     \\
J090958.07+621450.4 & 1.35                 & 2.94             & 0.011    & -    & -                         & 2.64 & G     \\
J081314.20+522731.4 & 1.24                 & 0.57             & 0.0      & -    & -                         & 2.34 & G     \\
J110047.95+104341.3 & 1.22                 & 0.73             & -0.003   & 9.95          & -0.91                              & 2.19 & G     \\
J080328.94+332744.6 & 1.18                 & 0.87             & -0.003   & 9.72          & -1.29                              & 3.27 & G     \\
J090022.63+645451.7 & 1.25                 & 1.3              & 0.003    & 10.2          & -0.54                              & 3.7  & G     \\
J161711.61+495751.9 & 1.24                 & 0.97             & 0.032    & -    & -                         & 2.93 & G     \\
J162415.16+201100.7 & 1.24                 & 1.47             & 0.025    & 10.32         & -0.53                              & 2.91 & G     \\
J141057.23+252950.0 & 1.21                 & 0.73             & 0.043    & 10.42         & -0.47                              & 2.14 & G     \\
J163038.87+350323.0 & 1.24                 & 2.06             & 0.018    & -    & -                         & 2.55 & G     \\
J094508.97+683540.4 & 1.25                 & 0.39             & 0.0      & 9.18          & -1.52                              & 2.17 & G     \\
J084141.20+403926.8 & 1.17                 & 0.78             & 0.016    & -    & -                         & 2.83 & G     \\
J125813.56+262536.7 & 1.2                  & 0.98             & -0.001   & -    & -                         & 2.37 & G     \\
J133130.19+333223.4 & 1.28                 & 0.69             & 0.012    & 9.86          & -0.8                               & 2.45 & G     \\
J105100.31+121714.5 & 1.29                 & 1.16             & 0.008    & 10.1          & -0.71                              & 2.28 & G     \\
J113914.88+170837.1 & 1.25                 & 0.53             & 0.014    & 10.48         & -0.04                              & 4.4  & G     \\
J090649.87+484620.9 & 1.25                 & 1.46             & 0.016    & -    & -                         & 2.33 & G     \\
J155801.84+145748.8 & 1.27                 & 0.94             & 0.014    & 10.3          & -0.31                              & 2.87 & G     \\
J154812.64+111646.2 & 1.27                 & 2.16             & 0.013    & 9.73          & -0.85                              & 2.51 & G     \\
J115122.68+532639.0 & 1.23                 & 0.86             & 0.004    & -    & -                         & 3.0  & G    
\label{tab:parameter_value_random}
\end{longtable}

\end{document}